\documentclass[aps,prb,preprint,showpacs,preprintnumbers,amsmath,amssymb]{revtex4}
\usepackage{amscd}
\usepackage{mathrsfs}
\usepackage{amsfonts}
\usepackage{amssymb}
\usepackage{amsmath}

\usepackage{graphicx}
\usepackage{dcolumn}
\usepackage{bm}
\newcommand{\mycolor}[0]{ (Color online) }

\begin{document}
\title{Optimal effective current operator for flux qubit accounting for inductive effects}
\author{Zheng Li$^{1,2}$, Tao Wu$^{1,3}$, Jianshe Liu$^{1,3}$}
\affiliation{${}^1$Tsinghua National Laboratory for Information Science and
  Technology, Beijing 100084, China\\${}^2$Department of Electronic
  Engineering, Tsinghua University, Beijing 100084, China\\
  ${}^3$Institute of Microelectronics, Tsinghua University, Beijing 100084,
  China} \email{lizheng02@mails.tsinghua.edu.cn}
\date{\today}

\begin{abstract} An optimal effective current operator for flux qubit has been
    investigated by taking account of the inductive effects of the circuit
    loop. The whole system is treated as two interacting subsystems: one is
    the inductance-free flux qubit consisting of three Josephson junctions
    and the other a high frequency LC-oscillator. As the
    composite system hardly affords one excessively high energy LC photon%
    ,
    an effective theory for the inductive flux qubit providing its
    physical variable operators has been achieved, which can take account
    of the inductive effects but does not include the additional degree of
    freedom for the LC-oscillator. Considering the trade-off between simplicity and accuracy, it has been revealed that the optimal effective
    current operator resulting in an error only on the order of $L^{3/2}$
    provides an approximation of high accuracy, which is also verified
    numerically.

\end{abstract}

\maketitle

\section{Introduction}

Superconducting circuits are promising candidates for quantum information
processing\cite{YGA2001,MAJ2004,Wendin2005} and, in order to reduce the impact of
both charge and flux noise, flux qubit consisting of a superconducting loop
interrupted by three Josephson junctions (3jj) has been proposed, designed and
realized.\cite{TJL1999, JTL1999, CAF2000, IYC2003, IPK2004, WYJ2005,JPC2007,DMB2008}
The loop in the original design is small enough and its inductive effects,
therefore, could be neglected at the first approximation.\cite{TJL1999} The
constraint of the flux quantization on the three phases across the 3jj yields
two independent phase variables for the system. On
the other hand, inductive effects are essential in several inductive coupling
schemes \cite{JFA2005, JYF2005, Alec2005, MG2006, YLiu2006, YLiu2007}. These
systems can be systematically studied by applying a general network graph
theory \cite{GRD2004,GDP2005}, which indicates that an independent phase is
associated to the loop self-inductance in the original circuit and the 3jj
flux qubit, thus, turns out as a three-phase system.  In order to include the
inductive effects judiciously, appropriate terms could be reallocated to
improve the original operators in the two-phase system. First, the inductive
effects, considered as corrections to the energy levels of the two-phase system,
have been addressed but with some flaws by Crankshaw et. al. in a semi-classical
approach\cite{DT2001}, and, consequently, an effective Hamiltonian has been
reached\cite{Alec2005} as well as a current operator in the two-flux-state basis
for the flux qubit\cite{Ya2006}. Another reason why we build up an
effective theory to include the inductive effects is that an
inductance of a non-negligible size may lead the device to a less useful
qubit.\cite{Robertson2006}

Current operator is crucial to the accurate control, coupling and measurement
of flux qubits.\cite{Lin2002} In particular, it could play a key role in
understanding the dynamics of the flux qubit by a general multilevel
model.\cite{WYJ2005,YJL2005,YWJ2005} Although various forms of current
operators have been utilized in all kinds of regimes, the validity of the
specific current operators has not been justified seriously and error analyses
are hardly available. In this work, a systematic investigation on the optimal
two-phase effective current operator for the three-phase system is carried out and an
error analysis is provided.

The paper is organized as follows. In Sec.\ref{sec:cl}, we review some basic ideas on the loop current in a classical circuit model. In Sec.\ref{sec:qu},
we construct the three-phase Hamiltonian and decompose it into a form showing that
two subsystems weakly interact with each other; then we develop an
effective theory, the photon transition path (PTP) approach based on the Brillouin-Wigner expansion\cite{ JJ1998}, to describe the three-phase system  in Sec.\ref{sec:ph}.  In Sec.\ref{sec:ef}, we obtain the optimal
effective two-phase loop current operator from the unique one for the whole system, and a brief numerical discussion is presented in
Sec.\ref{sec:br}.

\section{classical analysis}
\label{sec:cl}
\begin{figure}
\begin{minipage}[c]{0.45\textwidth}
\begin{center}
\includegraphics[keepaspectratio=true,width=.95\textwidth]{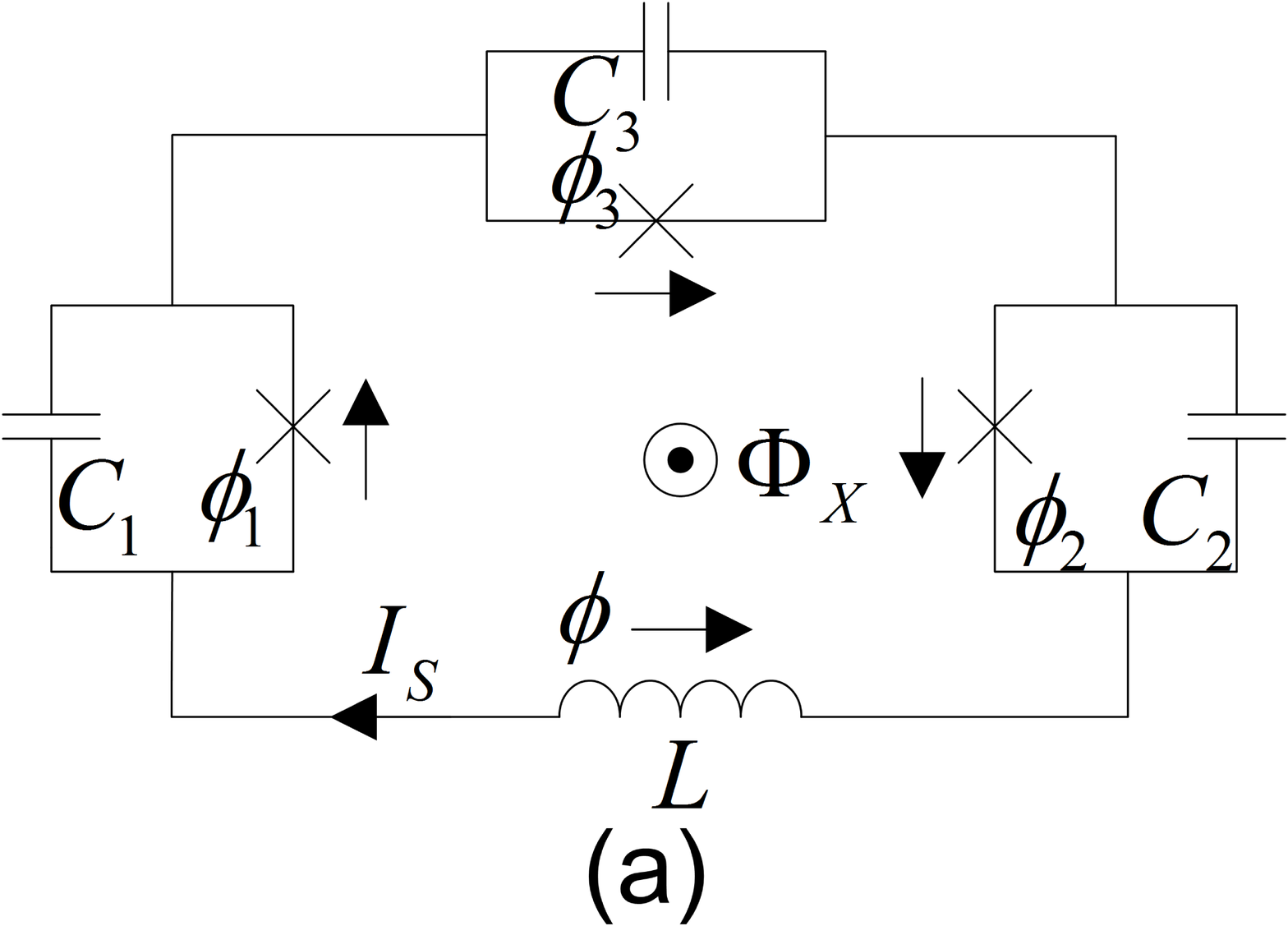}
\end{center}
\end{minipage}%
\begin{minipage}[c]{0.5\textwidth}
\begin{center}
\includegraphics[keepaspectratio=true,width=.95\textwidth]{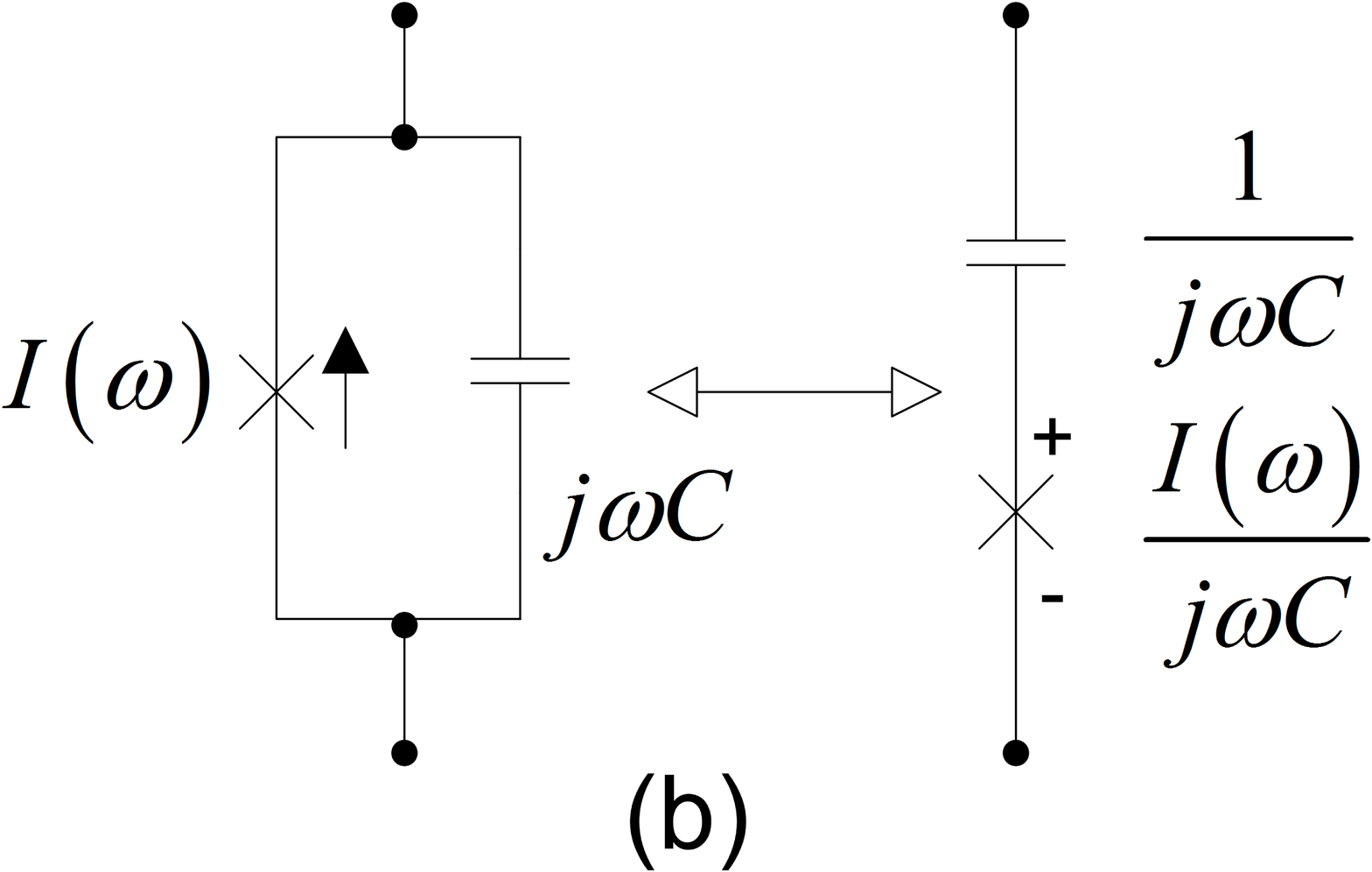}
\end{center}
\end{minipage}
\caption{(a)circuit of an inductive flux qubit with the phase
  difference $\phi$ across the loop inductance $L$, the reduced applied
  external flux $\phi_X={2\pi\Phi_X}/{\Phi_0}$ with $\Phi_0$ the flux %
  quantum, and the phase difference $\phi_k$ across the $k$th junction
  characterized via the critical current $I_{Ck}$ and the capacitance
  $C_k$ for $k$=1,2 and 3; (b) transformation between the current and
  voltage sources, the arrow and the plus/minus symbols indicate the
  directions of the current and voltage sources, respectively.}\label{fig:circuit}
\end{figure}
The schematic circuit for the 3jj flux qubit with a loop inductance
is demonstrated in Fig. \ref{fig:circuit}(a), where the 3rd junction is a
little smaller than those two others; representing the relative sizes,
the parameter $\alpha_k$ as
\begin{eqnarray}
  \alpha_k&=&\frac{I_{Ck}}{I_{C0}}=\frac{C_k}{C_0},k=1,2,3
\end{eqnarray}
indicates the area factor of the $k$th junction, where
$I_{C0}=\left(I_{C1}+I_{C2}\right)/2$ and $C_{0}=\left(C_{1}+C_{2}\right)/2$
are design parameters. Parameters $\alpha_1$ and $\alpha_2$ are supposed to be
close to 1, the deviations of which are determined by the accuracy of fabrication,
while $\alpha_3\simeq0.8$ and the reduced applied external flux $\phi_X$ is biased on the vicinity of
$\phi_X=\pi$, all of which are selected to benefit the
energy levels of the flux qubit. These three junction phases
$\phi_{1,2,3}$ and the phase difference $\phi$ across the loop inductance $L$ are not independent of one another and obey the flux quantization in this superconducting system as
\begin{eqnarray}
  \label{eq:flux_quan}
  \phi_1+\phi_2+\phi_3=\phi+\phi_X,
\end{eqnarray}
the signs of which are also indicated in Fig. \ref{fig:circuit}(a).

In the classical regime, the junction performs as a current-flux
2-port circuit element, a nonlinear inductance, and the flux
quantization condition imposes a predetermined constraint. In the DC
regime, without considering capacitances, the loop current flows
equivalently through four current elements in the loop including
the Josephson junctions and the loop inductance as
\begin{eqnarray}
  \label{eq:DC_current}
  I_{q}=I_{C_k}\sin\bar\phi_k=-\frac{\Phi_0}{2\pi}\frac{\bar\phi}{L}, k=1,2,3,
\end{eqnarray}
where $\bar\phi_1$, $\bar\phi_2$, $\bar\phi_3$ and $\bar\phi$ are
the possible static phase values obtained from Eqs. (\ref{eq:flux_quan}) and
(\ref{eq:DC_current}). Two opposite current directions present
an additional degeneracy of the circuit. Furthermore, in the AC regime, if only
taking account of the small oscillations in the circuit, each junction
works at the static phase point as a pure inductance
\begin{eqnarray}
  \label{eq:AC_LK}
  L_k=\frac{\Phi_0}{2\pi I_{C_k}\cos\bar\phi_k}
\end{eqnarray}
if $\cos{\bar\phi_k}\neq0$ for k=1,2 and 3. The series impedance of
the circuit,
\begin{eqnarray}
  \label{eq:AC_impedance}
  \mathbf{Z}(\omega)=-i\omega L+\sum_{k=1}^{3}\frac{-i\omega L_k}{1-\omega^2 L_kC_k},
\end{eqnarray}
with $i$ the imaginary unit, provides its several characteristic
frequencies; especially, when the circuit works at
such an ultra-high frequency that the junction inductances can be
treated as open circuits, there exists only one significant oscillation
along the loop between its small inductance $L$ and series capacitance
$C_{ser}$ with a high characteristic
frequency
\begin{eqnarray}
  \label{eq:AC_omega0}
  \omega_{LC}=\sqrt{\frac{1}{LC_{ser}}},
\end{eqnarray}
where
$C_{ser}=(\sum_{k=1}^3C_k^{-1})^{-1}=\alpha_{ser}C_{0}$.

Generally, the nonlinear effects of the junctions generate
current components of new frequencies different from the external flux-driven
source's.  Only considering the output profiles of the junctions, we can still
apply this kind of specific current sources to the rest of the
circuit which obeys the linear superposition rules. Picking an arbitrary
frequency $\omega$ in the frequency domain and utilizing a source transformation
shown in Fig. \ref{fig:circuit}(b), we have
\begin{eqnarray}
  \label{eq:AC_source}
  I_{loop}(\omega)=\frac{\sum_{k=1}^3\frac{I_{Ck}(\omega)}{i\omega
      C_k}}{\sum_{k=1}^3\frac{1}{i\omega C_k}+i\omega L},
\end{eqnarray}
where $I_{Ck}(\omega)$ is obtained from $I_{Ck}\sin\phi_k$ via the
Fourier transform.  Interestingly, when $L$ is small enough to
neglect, $I_{loop}(\omega)$ in Eq. (\ref{eq:AC_source}) does not
depend on $\omega$ explicitly and we utilize the inverse Fourier
transform $\mathcal{F}^{-1}$ again as
\begin{eqnarray} \label{eq:AC_time}
  I_{loop}(t)=\mathcal{F}^{-1}\left({I_{loop}(\omega)|_{\omega^2L\rightarrow0}}\right)={C_{ser}}
  \sum_{k=1}^3\frac{I_{Ck}\sin\phi_k}{C_k},
\end{eqnarray}
where $\sum_{k=1}^{3}\phi_k=\phi_X$ since $\phi$ vanishes when
$L\rightarrow0$. This form of the loop current $I_{loop}(t)$
directly goes with the fact that the junctions connect to a
topological network consisting of linear circuit elements. Delightfully, $I_{loop}(t)$ in Eq.
(\ref{eq:AC_time}) is in exact agreement with the one for the two-phase
system derived in the quantum regime by Maassen van den Brink\cite{Alec2005}
and with our following effective one. This suggests
that quantum superconducting circuit analysis and design might benefit in elegant
ways from classical circuit theories and CAD tools.

\section{quantum analysis for system Hamiltonian}
\label{sec:qu}
To construct the Hamiltonian comfortably, we firstly select three
junction phases $\phi_{1,2,3}$ as the spatial variables and express
the system Hamiltonian in a sum of energy terms similar to
other superconducting loop circuits such as the RF-qubit and the SQUID-qubit
as
\begin{eqnarray}\label{eq:ham_org}
  \hat{H}_{3p}=\sum_{k=1}^{3}\left(\frac{\hat{Q}_{k}^{2}}{2C_{k}}-E_{Jk}\cos\hat\phi_{k}\right)
  +\left(\frac{\Phi_{0}}{2\pi}\right)^{2}\frac{\hat\phi^{2}}{2L},
\end{eqnarray}
where $\hat Q_{k}$ is the charge operator conjugated with the phase
$\hat\phi_{k}$, i.e., $\hat{Q}_k=-2ei\frac{\partial}{\partial\phi_k}$ or
$\left[\hat\phi_{k},\hat Q_{k}\right]=2ei$ with $e$ the electronic
charge; $E_{Jk}=\frac{\Phi_{0}I_{Ck}}{2\pi}$ is the Josephson energy
of the $k$th junction. The former sum in Eq.(\ref{eq:ham_org}) represents
the total energy of junctions including their charge and
Josephson energy and the latter term the loop inductive
energy. According to the design, the reduced inductance
size $\beta$,
\begin{eqnarray}
  \label{eq:beta}
  \beta&=&\frac{2\pi{L}I_{C0}}{\Phi_0},
\end{eqnarray}
is usually small enough that the
loop phase difference $\hat{\phi}$ behaves as a small variable with
its norm $\|\hat\phi\|$ tending to be equal to zero, while the loop
current still keeps finite due to the biased junctions.
Consequentially, its conjugate variable $\hat Q_{\phi}$, which we refer to
as $\left[\hat\phi, \hat{Q}_\phi\right]=2ei$, diverges on its norm
according to the Heisenberg uncertainty principle
$\|\hat\phi\|\bullet\|\hat Q_{\phi}\|\gtrsim{e}$. In the classical regime, a quadratic potential
$\left(\frac{\Phi_{0}}{2\pi}\right)^{2}\frac{\phi^{2}}{2L}$ means that there is a generalized restoring force
\begin{eqnarray}
  \label{eq:force_osc}
  \vec{F}_{osc}&=&\frac{\Phi_0}{4\pi L}\nabla_{\phi_1,\phi_2,\phi_3}\phi^2=\frac{\Phi_0\phi}{2\pi L}[1,1,1]^T
\end{eqnarray}
providing a non-parallel generalized
acceleration
\begin{eqnarray}
  \label{eq:acce_osc}
  \vec{a}_{osc}&=&\mathbf{C}_{diag}^{-1}\vec{F}_{osc}=\frac{\Phi_0\phi}{2\pi LC_0}\vec{r},
\end{eqnarray}
where $\mathbf{C}_{diag}$ is a diagonal matrix with its diagonal
elements being $C_1$,$C_2$ and $C_3$, and %
$$\vec{r}=[\frac{1}{\alpha_1},\frac{1}{\alpha_2},\frac{1}{\alpha_3}]^T.$$
In the quantum regime,
the deep quadratic potential explicitly in proportion
to $1/\beta$ is capable to bind up the quantum states of this
three dimensional system in the vicinity of a phase plane $\phi=0$,
where a fast vacuum fluctuation occurs along the unique
direction $\vec{r}$ parallel to the acceleration $\vec{a}_{osc}$.
Therefore, the original spatial variable set
($\phi_{1}$,$\phi_{2}$,$\phi_{3}$), although helpful in the
construction of the Hamiltonian, presents difficulties in handling the charge
operator $\hat{Q}_{\phi}$, which represents one of the most important quantum properties of the
three-phase system.

To solve the problem, we utilize a linear transformation to achieve another set
of coordinates ($\phi$,$\theta_{1}$,$\theta_{2}$) where besides $\phi$ the other two
coordinates are labeled via $\theta_1$ and $\theta_2$ and their
conjugates are $\hat Q_{\theta_1}$ and $\hat
Q_{\theta_2}$, respectively.  The linear transformation between these two
sets of coordinates is introduced via a matrix $\mathbf{A}$ defined as
$[\theta_1, \theta_2, \phi+\phi_x]^T = \mathbf{A} [\phi_1, \phi_2,
\phi_3]^T$, or equivalently as
\begin{eqnarray} \label{eq:Q_theta}
    \mathbf{\hat Q}_\Theta^{T}=\left[\hat Q_{\theta_1}, \hat Q_{\theta_2}, \hat Q_{\phi}\right] =\left[\hat Q_1, \hat Q_2, \hat Q_3\right] \mathbf{A}^{-1}.
\end{eqnarray}
Thus, the Hamiltonian $\hat{H}_{3p}$ changes to
\begin{eqnarray}
  \label{eq:Ham_A}
  \hat{H}_A=\frac{1}{2}\mathbf{\hat
  Q}_\Theta^T\mathbf{A}\mathbf{C}_{diag}^{-1}\mathbf{A}^{T}\mathbf{\hat Q}_\Theta+V(\theta_1,\theta_2,\phi),
\end{eqnarray}
where $V(\theta_1,\theta_2,\phi)$ is the potential in the new
framework. Since the charge operator $\hat Q_{\phi}$ tends to
diverge when $\beta\rightarrow0$, if the charge coupling coefficients in
$\mathbf{A}\mathbf{C}_{diag}^{-1}\mathbf{A}^{T}$ are assumed to be finite and independent of $\beta$,
a proper candidate for $\Theta$-subsystem on ($\theta_{1},\theta_{2}$)
should avoid any direct charge
coupling from the $\phi$-subsystem. It mathematically requires
that the directions of $\theta_1$ and $\theta_2$ in the original
coordinates should be perpendicular to the acceleration direction $\vec{r}$
of the oscillation mentioned above, which means
that the plane spanned by $\theta_1$ and $\theta_2$ is
unique as well as
\begin{eqnarray}
  \label{eq:Q_phase}
  \hat{Q}_\phi&=&{C_{ser}}\sum_{k=1}^{3}\frac{\hat{Q}_k}{C_k},
\end{eqnarray}
revealing the charge in the series capacitor $C_{ser}$. Some other explanations in the classical regime
are also given in Refs. \onlinecite{Alec2005} and \onlinecite{DT2001}, both of
which have achieved the proper variable transformations by avoiding the cross
charge energy terms between $\Theta$ and $\phi$ subsystems. %
\footnote{%
Scrutinizing the Kirchhoff equations in Ref. \onlinecite{Alec2005}
is of less difference from the classical discussion on
Hamiltonian in Ref. \onlinecite{DT2001}. No cross term in the
classical regime mathematically is equivalent to the corresponding quantum case due to the same
matrix $\mathbf{A}$ making both
$\left(\mathbf{A}\mathbf{C}_{diag}^{-1}\mathbf{A}^{T}\right)^{-1}$ (in the
classical regime) and $\mathbf{A}\mathbf{C}_{diag}^{-1}\mathbf{A}^{T}$ (in the
quantum regime) take $[0,0,1]^{T}$ as one of their eigenvectors.}%
Although they have predicted the right ones based on the
linearity of the circuit, it is more comfortable in the quantum regime to
emphasize the reason why the $\Theta$-subsystem as well as $\hat{Q}_{\phi}$ should be selected
uniquely, since
the diverging charge fluctuations merely serve as a pure quantum
phenomenon.

The remaining degrees of freedom endowed by $\mathbf{A}$ involve the
internal variable selections of $\Theta$-subsystem. A straightforward
way is that $\theta_1$ and $\theta_2$
only deviate slightly from $\phi_1$ and $\phi_2$, respectively; then,
the whole transformation reads as follows,
\begin{eqnarray}
\label{eq:theta_trans}
  \left[
    \begin{array}{c}
      \theta_1\\\theta_2\\\phi
    \end{array}
  \right]=
  \left[
    \begin{array}{ccc}
      1-\frac{C_{ser}}{C_1}&-\frac{C_{ser}}{C_1}&-\frac{C_{ser}}{C_1}\\
      -\frac{C_{ser}}{C_2}&1-\frac{C_{ser}}{C_2}&-\frac{C_{ser}}{C_2}\\
      1&1&1
    \end{array}
  \right]\left[
    \begin{array}{c}
      \phi_1\\\phi_2\\\phi_3
    \end{array}
  \right]+\left[
    \begin{array}{c}
      \frac{C_{ser}}{C_1}\\\frac{C_{ser}}{C_2}\\-1
    \end{array}
  \right]\phi_X,
\end{eqnarray}
where the last term of its right-side is a set of
constant biases as a translation in the superconducting phase space. For
short, it can also be reformatted as
\begin{equation}
  \label{eq:theta_for_short}
  \theta_{k}=\phi_{k}-\frac{C_{ser}}{C_{k}}\phi,k=1,2,
\end{equation}
which clearly shows that $\phi_{k}$ reduces to $\theta_{k}$ when $\phi\rightarrow0$.
The transformed charge operators
\begin{eqnarray}
  \label{eq:Q_trans}
  \left[
    \begin{array}{c}
      \hat Q_{\theta_1}\\\hat Q_{\theta_2}\\\hat Q_{\phi}
    \end{array}
  \right]=\left[
    \begin{array}{ccc}
      1&0&-1\\
      0&1&-1\\
      \frac{C_{ser}}{C_1}&\frac{C_{ser}}{C_2}&\frac{C_{ser}}{C_3}
    \end{array}
  \right]\left[
    \begin{array}{c}
      \hat Q_1\\\hat Q_2\\\hat Q_3
    \end{array}
  \right]
\end{eqnarray}
indicate that $\hat Q_{\theta_1}$ states the charge of the island
between the junctions 1 and 3, and analogously for $\hat Q_{\theta_2}$. If
we also define
\begin{eqnarray}
  \theta_{3}=\phi_{3}-\frac{C_{ser}\phi}{C_{3}},
  \label{eq:theta3}
\end{eqnarray}
equating to $\phi_{X}-\theta_{1}-\theta_{2}$, three new phase
variables $\theta_{1,2,3}$ confined by the flux quantization seem to
act as the junction phases in the two-phase system, which is confirmed by the following
transformed Hamiltonian
\begin{eqnarray}
\label{eq:H_tr_pre}
\hat{H}_{tr}&=&\hat{H}_{0}+\left(\hat{a}^{\dag}\hat{a}+\frac{1}{2}\right)\hbar\omega_{LC}+\hat{H}_{int},
\end{eqnarray}
where
\begin{eqnarray}
  \hat{H}_0&=&\frac{1}{2}{\hat{\mathbf{Q}}_{\theta}^{T}\mathbf{C}_{2p}^{-1}\hat{\mathbf{Q}}_\theta}-\sum_{k=1}^{3}E_{Jk}\cos\hat{\theta}_{k},
  \\\hat{a}^{\dagger}\hat{a}&=&\frac{1}{\hbar\omega_{LC}}\left(\frac{\hat{Q}^2_\phi}{2C_{ser}}+\left(\frac{\Phi_{0}}{2\pi}\right)^{2}\frac{\hat{\phi}^{2}}{2L}\right)-\frac{1}{2},
  \\\hat{H}_{int}&=&\sum_{k=1}^{3}E_{Jk}\cos\hat{\theta}_{k}-\sum_{k=1}^{3}E_{Jk}\cos\left(\hat{\theta}_{k}+\frac{C_{ser}}{C_k}\hat{\phi}\right),\\
  \hat{\mathbf{Q}}_\theta&=&\left[
    \begin{array}{c}
      \hat{Q}_{\theta_1}\\\hat{Q}_{\theta_2}
    \end{array}\right],\\
  \mathbf{C}_{2p}&=&\left[
    \begin{array}{cc}
      C_1+C_3   &   C_3\\
      C_3   &   C_2+C_3
    \end{array}\right].
\end{eqnarray}
The Hamiltonian has been decomposed into three parts. The first part $\hat
H_0$ is the Hamiltonian of the $\Theta$-subsystem for the inductance-free flux
qubit.\cite{TJL1999} The middle part
$(\hat{a}^{\dagger}\hat{a}+1/2)\hbar\omega_{LC}$ shows that the LC oscillator
consists of the phase variable ${\phi}$ and its conjugate $\hat Q_\phi$ and its
characteristic frequency $\omega_{LC}$=$1\slash\sqrt{LC_{ser}}$ is high enough
as mentioned in the classical regime. The operators $\hat a^{\dag}$ and $\hat a$ are defined
respectively as the photon creation and annihilation operators and the
dimensional factor of $\hat\phi$=$\frac{2\pi}{\Phi_{0}}\sqrt[4]{\frac{L\hbar^{2}}{C_{ser}}}\frac{\left(\hat
a^{\dag}+\hat a\right)}{\sqrt{2}}$ is in proportion to $\beta^{1/4}$. The
last part $\hat H_{int}$ is the interaction Hamiltonian between these two
subsystems, which is weak enough compared with
$\hbar\omega_{LC}\hat{a}^\dag\hat{a}$ and $\hat{H}_0$ to make the $\Theta$-subsystem only be
slightly perturbed by the LC oscillator. The current implementation of $\hat H_{int}$ is useful for
numerical solutions of diagonalizing Kronecker product matrices with the FFT
tools,\cite{TZJ2006} which are also utilized in this paper, and its
series expansion on $\hat{\phi}$
\begin{equation}
  \label{eq:H_int}
  \hat H_{int}=\sum_{k\geq1}\hat{V}_{k}\hat\phi^{k},
\end{equation}
where $\hat
V_{k}=\textstyle\frac{1}{k!}\frac{\partial^{k}}{\partial\phi^{k}}\hat{H}_{int}|
_{\phi=0}$, is fit for the perturbation methods which should deal with the couplings of different strengths.

\section{Photon transition path method}
\label{sec:ph}
\begin{figure}
\begin{minipage}[c]{0.45\textwidth}
\begin{center}
\includegraphics[keepaspectratio=true,width=1\textwidth]{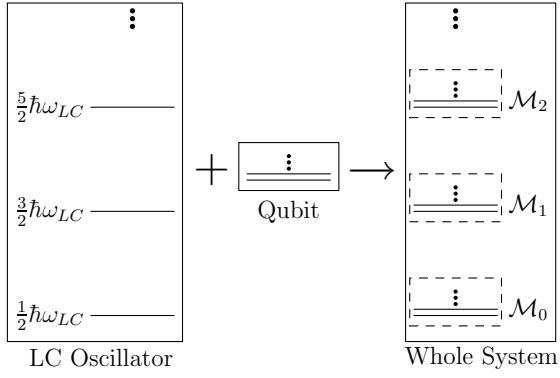}
\end{center}
\end{minipage}
\caption{Energy diagram of flux qubit with a loop inductance. When the inductance-free flux
qubit and the LC oscillator interact with each other in a perturbation
condition, the lowest eigenstates in the dressed state manifold $\mathcal{M}_{0}$
denoted with the dashed-line
box are well separated from the ones in other manifolds $\mathcal{M}_{1}$, $\mathcal{M}_{2}$, $\cdots$ due to the large shifting caused by the LC-photon energy $\hbar\omega_{LC}$.}
\label{fig:dressed}
\end{figure}
\begin{figure}
\begin{minipage}[c]{0.45\textwidth}
\begin{center}
\includegraphics[keepaspectratio=true,width=.85\textwidth]{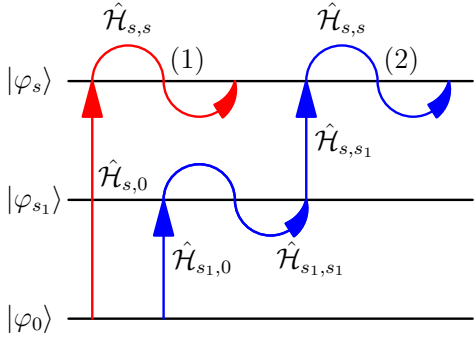}
\end{center}
\end{minipage}
\caption{\mycolor{}Two typical photon transition paths (1) and (2) represented by the
linked operator chains $\hat{\mathcal H}_{s,s}(\varepsilon)\hat{\mathcal
H}_{s,0}$ and $\hat{\mathcal H}_{s,s}(\varepsilon)\hat{\mathcal
H}_{s,s_{1}}\hat{\mathcal H}_{s_{1},s_{1}}(\varepsilon)\hat{\mathcal
H}_{s_{1},0}$, respectively.}
\label{fig:chain}
\end{figure}

\subsection{Dressed states in manifolds}
\label{sec:ph:dr}

To understand the energy diagram of this system shown in Fig.
\ref{fig:dressed}, let us briefly recapitulate the well-known dressed state
concept\cite{Atom1992}. %
For the sake of simplicity, we do not explicitly consider the possible inner degeneracy in
the two-phase subsystem further and have a set of complete orthogonal basis
$\mathcal{B}_{\hat{H}_{0}}=\left\{|\varphi_{f}\rangle\left|\hat{H}_{0}|\varphi_{f}\rangle=\varepsilon_{f}|\varphi_{f}\rangle\right.\right\}$
, where the arbitrary normalized eigenstate $|\varphi_{f}\rangle$ goes with its
eigenenergy $\varepsilon_{f}$. The LC-oscillator keeps its $s$th
eigenstate $|\Omega_s\rangle$ as $\hat{a}^\dag\hat{a}|\Omega_s\rangle =
s|\Omega_s\rangle$.

When the interaction $\hat{H}_{int}$ is neglected at the first approximation,
it is convenient to find that a series of dressed artificial-atom states
$\{|\varphi_{f}\rangle\otimes|\Omega_{0}\rangle,~\cdots,~|\varphi_{f}\rangle\otimes|\Omega_{N}\rangle,~\cdots\}$,
where $N$ is a non-negative integer, are the eigenstates of the whole
system with their eigenenergy being
$\{\varepsilon_{f}+\frac{1}{2}\hbar\omega_{LC},~\cdots,~\varepsilon_{f}+(N+\frac{1}{2})\hbar\omega_{LC},~\cdots\}$,
respectively.  Since $\hbar\omega_{LC}\propto\beta^{-1/2}$ is
much larger than $\varepsilon_{f}$ independent of $\beta$,
the dressed states are so well separated from one another that the
tensor-product states which keep the same photon number can be grouped together
to form one so-called manifold. For example, the $N$th manifold
$\mathcal{M}_{N}^{0}=\left\{|\varphi\rangle\left||\varphi_{f}\rangle\otimes|\Omega_{N}\rangle,\forall|\varphi_{f}\rangle\in\mathcal{B}_{\hat{H}_{0}}\right\}\right.$
consists of all possible eigenstates possessing $N$ LC photons and maintains
the same energy level structure as the two-phase flux qubit's if $(N+\frac{1}{2})\hbar\omega_{LC}$ is subtracted. %
After the weak interaction $\hat{H}_{int}$ turns on in the order analysis, the possible intra- and
inter-manifold photon-assisted transitions bring perturbations of different strengths, which cannot completely destroy the manifold structures, so the perturbed
eigenstates in the $N$th manifold $\mathcal{M}_{N}$ can still be distinguished
from other manifolds' due to the $(N+\frac{1}{2})\hbar\omega_{LC}$
energy shifting. This kind of understanding can be
revealed by one well-known perturbation approach, the unitary transformation
(UT) method\cite{Atom1992}, which introduces a specific unitary transformation
$\hat{T}=e^{i\hat{S}}|_{\hat{S}=\hat{S}^\dag}$, resembling a time-evolution operator, to rotate the Hamiltonian $\hat{H}_{tr}$
into a new one $\hat{H}^{'}_{tr}=\hat{T}^{\dag}\hat{H}_{tr}\hat{T}$ so that it
can be diagonalized as
$\hat{H}'_{tr}=\sum_{N=0}^{\infty}\tilde{{H}}_{N}|\Omega_{N}\rangle\langle\Omega_{N}|$
on an arbitrary order of $\beta$. The two-phase Hamiltonian
$\tilde{{H}}_{N}$ performs as an effective one for the $N$th manifold
$\mathcal{M}_{N}$: with the eigenstate basis
$\mathcal{B}_{\tilde{{H}}_{N}}=\left\{|\tilde{\varphi}_{N}\rangle\left|\tilde{H}_{N}|\tilde{\varphi}_{N}\rangle=\tilde{\varepsilon}_{N}|\tilde{\varphi}_{N}\rangle\right.\right\}$, %
the $N$th manifold $\mathcal{M}_{N}$ can be rewritten as
$\mathcal{M}_{N}=\left\{|\varphi\rangle\left||\varphi\rangle=\hat{T}\left(|\tilde{\varphi}_{N}\rangle\otimes|\Omega_{N}\rangle\right),\forall|\tilde{\varphi}_{N}\rangle\in\mathcal{B}_{\tilde{H}_{N}}\right\}\right.$.
In particular, when $\hat{H}_{int}$ is neglected we can select
$\hat{S}=0$ and obtain the effective Hamiltonian
$\tilde{{H}}_{N}^{(0)}=(\frac{1}{2}+N)\hbar\omega_{LC}+\hat{H}_{0}$ indicating
that the manifold $\mathcal{M}_{N}$ unsurprisingly becomes
$\mathcal{M}_{N}^{(0)}$ when the interactions turn off. With the unitary
operator $\hat{T}$, one can also consequently construct other effective
operators.

On the other hand, since there is hardly an experimental way to keep the
high-energy LC-oscillator excited in the superconducting circuit applications, what
needs to be focused on actually is the lowest eigenstates belonging to the
manifold $\mathcal{M}_{0}$. This physical requirement also enables us to
circumvent the additional discussions on that the
inductance-free flux qubit as an infinite-level system still leaves the
high-energy eigenstates in $\mathcal{M}_{0}$ not being well separated from but
overlapping with the lowest ones in $\mathcal{M}_{1}$ for a specific value $\beta$ in the energy diagram.
By means of the Rayleigh-Schr$\ddot{\mbox{o}}$dinger (RS) expansion
with the arbitrary eigenstate
$|\varphi\rangle$ in $\mathcal{M}_{0}$ and its eigenenergy $\varepsilon$
being respectively expanded as
\begin{eqnarray}
\label{eq:state_exp}
|\varphi\rangle&=&
|\varphi^{(0)}\rangle+|\varphi^{(1)}\rangle+|\varphi^{(2)}\rangle+\dots,\\
\varepsilon&=&
\frac{1}{2}\hbar\omega_{LC}+\varepsilon^{(0)}+\varepsilon^{(1)}+\varepsilon^{(2)}+\dots,\label{eq:eps_exp}
\end{eqnarray}
where both $|\varphi^{(k)}\rangle$ and $\varepsilon^{(k)}$ are in proportion to
$\beta^{k/4}$ for $k$ as an integer and $|\varphi^{(0)}\rangle$ belongs to
the manifold $\mathcal{M}_{0}^{(0)}$,  an effective Hamiltonian
on the order of $\beta$ has been obtained but without further discussions on the
higher order expansions in Ref.  \onlinecite{Alec2005}. In this paper, based on the
Brillouin-Wigner(BW) expansion, another famous perturbation
theory, we develop a photon transition path method to further explore the perturbation procedure %
and compare it with the one in Ref.  \onlinecite{Alec2005} and also with the
UT method. %

\subsection{Formal substitution derivation}

Besides the order expansion in Eq.
(\ref{eq:state_exp}), we also expand $|\varphi\rangle$ in the energy eigenbasis of the
oscillator as
\begin{eqnarray}
  \label{eq:phi_exp}
  |\varphi\rangle&=&\sum_{s=0}^\infty|\varphi_s\rangle\otimes|\Omega_s\rangle=\sum_{s=0}^\infty\sum_{k=0}^\infty|\varphi_s^{(k)}\rangle\otimes|\Omega_s\rangle,
\end{eqnarray}
where $|\varphi_s\rangle=\langle\Omega_s|\varphi\rangle$, and
$|\varphi_s^{(k)}\rangle\propto\beta^{k/4}$. Since the expansion in Eq. (\ref{eq:eps_exp}) begins with a constant number $\frac{1}{2}\hbar\omega_{LC}$, we subtract
it from $\hat{H}_{tr}$ and redefine the Hamiltonian
$\hat{H}_{tr}$ as
\begin{eqnarray} \label{eq:H_tr}
    \hat{H}_{tr}&=&\hat{H}_{0}+\hbar\omega_{LC}\hat{a}^{\dag}\hat{a}+\hat{H}_{int}.
\end{eqnarray}
Consequently, the biased eigenenergy $\varepsilon$ satisfies $\varepsilon\ll\hbar\omega_{LC}$. The
Hamiltonian $\hat{H}_{tr}$ can be expanded as a Kronecker product matrix
\begin{eqnarray}
  \label{eq:Ham_mat}
  \hat{H}_{tr}=\left[
    \begin{array}{ccccc}
      \hat{H}_{0,0}&\hat{H}_{0,1}&\cdots&\hat{H}_{0,s}&\cdots\\
      \hat{H}_{1,0}&\hat{H}_{1,1}+\hbar\omega_{LC}&\cdots&\hat{H}_{1,s}&\cdots\\
      \vdots&\vdots&\ddots&\vdots&\cdots\\
      \hat{H}_{s,0}&\hat{H}_{s,1}&\vdots&\hat{H}_{s,s}+s\hbar\omega_{LC}&\cdots\\
      \vdots&\vdots&\vdots&\vdots&\ddots
    \end{array}
    \right],
\end{eqnarray}
where the operator
$\hat{H}_{s,s}=\hat{H}_{0}+\langle\Omega_s|\hat{H}_{int}|\Omega_s\rangle$
refers to the self-transition of the $s$th level and
$\hat{H}_{s,s_{1}}=\langle\Omega_s|\hat{H}_{int}|\Omega_{s_{1}}\rangle=\hat{H}_{s_{1},s}^{\dag}$
the transition between the $s$th and $s_{1}$th levels of the oscillator. %
The operator $\hat{H}_{s,s}$ consists of terms of different strengths due to the
nonlinearities included in $\hat{H}_{int}$, and its dominant term $\hat{H}_{0}$ approximating $\hat{H}_{s,s}$ on $O(\beta^{0})$
suffers from an error on $O(\beta^{1/2})$ instead of $O(\beta^{1/4})$ thanks to
the optical selection rules. %
For the same reason, the dominant terms of the
operators $\hat{H}_{s,s_{1}}$ are also weakened on $O(\beta^{|s-s_1|/4})$.
More details about the order discussion are presented in Appendixes
\ref{sec:order} and \ref{sec:op_exp}.
The eigen-equation $\hat{H}_{tr}|\varphi\rangle=\varepsilon|\varphi\rangle$ is decomposed into a series of equations as
\begin{eqnarray}
  \left(\hat{H}_{0,0}-\varepsilon\right)|\varphi_0\rangle&=&-\sum_{k\ne{0}}\hat{H}_{0,k}|\varphi_k\rangle,\label{eq:eigen_exp_0} \\
  \left(s\hbar\omega_{LC}+\hat{H}_{s,s}-\varepsilon\right)|\varphi_s\rangle&=&-\sum_{k\ne{s}}\hat{H}_{s,k}|\varphi_k\rangle~(s>0). \label{eq:eigen_exp_s}
\end{eqnarray}
The shifting of $\frac{1}{2}\hbar\omega_{LC}$, thus, distinguishes Eq. (\ref{eq:eigen_exp_0})
from the others in Eq. (\ref{eq:eigen_exp_s}). It is clear that when the
loop inductive effects are totally neglected, this equation is capable to degrade into a two-phase eigen-problem as
\begin{eqnarray}
    \hat{H}_{0}|\varphi_{f}\rangle=\varepsilon_{f}|\varphi_{f}\rangle,
\end{eqnarray}
and all of the other projected states
$|\varphi_s^{(0)}\rangle_{(s>0)}$ equate to zero, which suggests that the
$\Theta$-subsystem decoupled from the LC-oscillator becomes an
inductance-free two-phase system and there is no LC-photon excited at the first approximation.

Since the LC-oscillator is of high energy,
$s\hbar\omega_{LC}$ always dominates in Eq.(\ref{eq:eigen_exp_s}) at the
excited levels (we assume that the integer $s$ is larger than zero in the
following sections). To figure out the relative strength, a new set of
operators are defined as
\begin{eqnarray}
\hat{\mathcal H}_{s,k}\overset{\Delta}{=}\frac{\varepsilon\delta_{s,k}-\hat H_{s,k}}{s\hbar\omega_{LC}}
\end{eqnarray}
with an introduced
Kronecker delta function $\delta_{a,b}$, thus Eq. (\ref{eq:eigen_exp_s})
yielding
\begin{eqnarray}
  \label{eq:eigen_exp_1}
  \left(1-\hat{\mathcal
  H}_{s,s}(\varepsilon)\right)|\varphi_s\rangle&=&\sum_{k\ne{s}}\hat{\mathcal H}_{s,k}|\varphi_k\rangle.
\end{eqnarray}
As the dominant term of $\hat{\mathcal H}_{s,s}(\varepsilon)$ is on $O(\beta^{1/2})$, we
expand the modified BW resolvent operator $(1-\hat{\mathcal
H}_{s,s}(\varepsilon))^{-1}$ as $\sum_{i\ge0}\hat{\mathcal
H}_{s,s}^{i}(\varepsilon)$, and  the state
  $|\varphi_{s}\rangle$ is given as
\begin{equation}
  |\varphi_{s}\rangle=\sum_{i\ge0}\sum_{s_{1}\neq s}\hat{\mathcal
    H}_{s,s}^{i}(\varepsilon)\hat{\mathcal
    H}_{s,s_{1}}|\varphi_{s_{1}}\rangle.\label{eff:s_solved}
\end{equation}
This equation indicates that $|\varphi_{s}\rangle$, different from
$|\varphi_{0}\rangle$, is a result of transitions from all of the other levels. Moreover,
it is found that the large photon energy  $s\hbar\omega_{LC}$ and the
corresponding multi-photon processes involved punish all of this kind of
transitions via diminishing them on specific orders of $\beta$.

According to the difference of the state $|\varphi_0\rangle$ from the others,
we leave the right hand side of Eq. (\ref{eff:s_solved}) separated as
\begin{equation}
  \label{eff:sep}
  |\varphi_{s}\rangle=\sum_{i\ge0}\hat{\mathcal
    H}_{s,s}^{i}(\varepsilon)\hat{\mathcal
    H}_{s,0}|\varphi_{0}\rangle+\sum_{i\ge0}\sum_{s_{1}\neq
    \{s,0\}}\hat{\mathcal H}_{s,s}^{i}(\varepsilon)\hat{\mathcal
    H}_{s,s_{1}}|\varphi_{s_{1}}\rangle,
\end{equation}
which involves two types of PTPs to the $s$th level: the operator $\hat{\mathcal
H}_{s,s}^{i}(\varepsilon)\hat{\mathcal H}_{s,0}$ means that the
state $|\varphi_0\rangle$ transfers from the ground level, then
through arbitrary times of self-transitions, to the $s$th level and
$\hat{\mathcal H}_{s,s}^{i}(\varepsilon)\hat{\mathcal H}_{s,s_{1}}$
refers to the other state $|\varphi_{s_1}\rangle$ (neither $|\varphi_{0}\rangle$ nor $|\varphi_{s}\rangle$) from the $s_{1}$th level. Since
$s_{1}\neq0$ in the above sum, we also have
\begin{equation}
  \label{eff:sep_s1}
  |\varphi_{s_{1}}\rangle=\sum_{i_{1}\ge0}\hat{\mathcal
    H}_{s_{1},s_{1}}^{i_{1}}(\varepsilon)\hat{\mathcal
    H}_{s_{1},0}|\varphi_{0}\rangle+\sum_{i_{1}\ge0}\sum_{s_{2}\neq
    \{s_{1},0\}}\hat{\mathcal H}_{s_{1},s_{1}}^{i_{1}}(\varepsilon)\hat{\mathcal
    H}_{s_{1},s_{2}}|\varphi_{s_{2}}\rangle,
\end{equation}
where $s$, $s_{1}$ and $i$ have been substituted by $s_{1}$, $s_{2}$
and $i_{1}$, respectively. The latter type of PTPs in Eq.(\ref{eff:sep}), therefore, can also be divided again as
\begin{eqnarray}
  \label{eff:long}
  |\varphi_{s}\rangle&=& \sum_{i\ge0}\hat{\mathcal
    H}_{s,s}^{i}(\varepsilon)\hat{\mathcal
    H}_{s,0}|\varphi_{0}\rangle+\sum_{i,i_{1}}\sum_{s_{1}\neq\{0,s\}}\hat{\mathcal
    H}_{s,s}^{i}(\varepsilon)\hat{\mathcal H}_{s,s_{1}}\hat{\mathcal
    H}_{s_{1},s_{1}}^{i_{1}}(\varepsilon)\hat{\mathcal
    H}_{s_{1},0}|\varphi_{0}\rangle+\\\nonumber&&\sum_{i,i_{1}}\sum_{s_{1}\neq\{0,s\}}\sum_{s_2\ne\{0,s_1\}}\hat{\mathcal
    H}_{s,s}^{i}(\varepsilon)\hat{\mathcal H}_{s,s_{1}}\hat{\mathcal
    H}_{s_{1},s_{1}}^{i_{1}}(\varepsilon)\hat{\mathcal
    H}_{s_{1},s_2}|\varphi_{s_2}\rangle.
\end{eqnarray}
These substitutions employ the procedures of the BW perturbation approach with a clearer view on the orders of the terms on $\beta$. For
example, without the expansions of the BW resolvent, Eq. (\ref{eff:long})
resembles a familiar BW expansion as
\begin{eqnarray}
  \label{eff:long:BW}
  |\varphi_{s}\rangle&=& \frac{1}{\varepsilon-s\hbar\omega_{LC}-\hat{H}_{s,s}}\hat{H}_{s,0}|\varphi_{0}\rangle+\\\nonumber&&
  \sum_{s_{1}\neq\{0,s\}}\frac{1}{\varepsilon-s\hbar\omega_{LC}-\hat{H}_{s,s}}\hat{H}_{s,s_{1}}\frac{1}{\varepsilon-s_{1}\hbar\omega_{LC}-\hat{H}_{s_{1},s_{1}}}\hat{H}_{s_{1},0}|\varphi_{0}\rangle+\\\nonumber&&
\sum_{s_{1}\neq\{0,s\}}\sum_{s_2\ne\{0,s_1\}}\frac{1}{\varepsilon-s\hbar\omega_{LC}-\hat{H}_{s,s}}\hat{H}_{s,s_{1}}\frac{1}{\varepsilon-s_{1}\hbar\omega_{LC}-\hat{H}_{s_{1},s_{1}}}\hat{H}_{s_{1},s_{2}}|\varphi_{s_{2}}\rangle.
\end{eqnarray} %
Furthermore, we are also able to substitute the $3$rd part in Eq. (\ref{eff:long}) and divide it into
two parts, the latter one of which can be substituted again. %
After retaining the transition paths to the $s$th level from the ground level
and continuing this kind of substitutions for $n-2$ times with $s_3$, $s_4$,
$\cdots$, and $s_{n}$ being introduced, we transform Eq. (\ref{eff:sep}) into
\begin{eqnarray}
  \label{eq:generator}
  |\varphi_s\rangle=\hat{G}^{(n)}_s(\varepsilon)|\varphi_0\rangle+\sum_{s_{n}\ne0}\hat{G}^{(n)}_{s,s_{n}}(\varepsilon)|\varphi_{s_{n}}\rangle,
\end{eqnarray}
where $\hat{G}^{(n)}_s(\varepsilon)$ refers to all of the PTPs
from the ground level involving \textit{no more than} $n$ times of
non-self transitions and $\hat{G}^{(n)}_{s,s_{n}}(\varepsilon)$
all of the PTPs in a form like $\hat{\mathcal
H}_{s,s}^{i}(\varepsilon)\hat{\mathcal H}_{s,s_{1}}\hat{\mathcal
H}_{s_{1},s_{1}}^{i_{1}}(\varepsilon)\hat{\mathcal
H}_{s_{1},s_2}\cdots\hat{\mathcal
H}_{s_{n-1},s_{n-1}}^{i_{n-1}}(\varepsilon)\hat{\mathcal H}_{s_{n-1},s_{n}}$,
the dominant terms of which are at least on $O(\beta^{n/2})$ contributed
by $\hbar\omega_{LC}$. Since $n$ can increase so large as to make
$\hat{G}^{(n)}_{s,s_{n}}(\varepsilon)$ negligible on an arbitrary order of $\beta$, the $s$th level is uniquely
determined by the state $|\varphi_{0}\rangle$ with a corresponding operator
$\hat{\mathcal G}_{s}(\varepsilon)$ being defined by
\begin{eqnarray}
  \label{eq:G_eps_def}
  |\varphi_{s}\rangle\overset{\Delta}{=}\hat{\mathcal
  G}_{s}(\varepsilon)|\varphi_{0}\rangle,
\end{eqnarray}
where $\hat{\mathcal G}_0$ can also be added as an identity operator
$\hat{{I}}_{2p}$ with $\hat{{I}}_{2p}|\varphi_0\rangle=|\varphi_0\rangle$.
For example, the operators $\hat{\mathcal G}_{1,2,3}(\varepsilon)$ are
approximately given in Appendix \ref{sec:op_exp}. %
Therefore, the projected state $|\varphi_0\rangle$ with a \textit{map}, which
a series of operators $\hat{\mathcal G}_{s}(\varepsilon)$ function as, covers the
three-phase state $|\varphi\rangle$ completely and accurately. %
Equation (\ref{eq:G_eps_def}) mathematically describes one physical understanding that for the states in the manifold
$\mathcal{M}_{0}$ photons persisting in all of the excited levels
come from the ground level via all possible PTPs as illustrated in Fig.
\ref{fig:chain} due to the
perturbations of the inductance-free flux-qubit.

Substituting $|\varphi_{s}\rangle$ in Eq. (\ref{eq:eigen_exp_0}) with the aid of
Eq. (\ref{eq:G_eps_def}), we have an eigen-like problem
\begin{equation}
  \label{eq:eff:pseudo}
  \tilde{H}(\varepsilon)|\varphi_{0}\rangle=\varepsilon|\varphi_{0}\rangle,
\end{equation}
where the \textit{pseudo}-Hamiltonian $\tilde{H}(\varepsilon)$ is defined as
\begin{equation}
  \label{eq:eff:pseudo:op}
  \tilde{H}(\varepsilon)=\hat{H}_{0,0}+\sum_{s>0}\hat{H}_{0,s}\hat{\mathcal{G}}_{s}(\varepsilon).
\end{equation}
In the definition of $\tilde{H}(\varepsilon)$, %
all of the terms in the latter sum can
be described in a general form $\hat{H}_{0,s}\hat{\mathcal
H}_{s,s}^{i}(\varepsilon)\hat{\mathcal H}_{s,s_{1}}\hat{\mathcal
H}_{s_{1},s_{1}}^{i_{1}}(\varepsilon)\hat{\mathcal
H}_{s_{1},s_2}\cdots\hat{\mathcal
H}_{s_{n-1},s_{n-1}}^{i_{n-1}}(\varepsilon)\hat{\mathcal H}_{s_{n-1},0}$ which can be interpreted
in the terms of the photon-assisted transitions as that the LC photons spread to one specific excited level such as the $s$th
one from the ground level through an arbitrary PTP (the role
$\hat{\mathcal{G}}_{s}(\varepsilon)$ plays) and then return back ( an operator
$\hat{H}_{0,s}$ closes the whole PTP). Therefore, the PTPs introduced by the
operator $\tilde{H}(\varepsilon)$ are not only linked but also closed, starting
from and ending with the ground level. It should be emphasized that a
one-to-one correspondence is established between the terms in this sum and the
closed photon transition paths(CPTP). Putting aside the details of the CPTPs in this section, one idea can be
accepted that the longer path the photons travel along, the weaker effects are
brought. %
Based on the BW expansion, the above derivations do not lose any accuracy thanks
to the \textit{formal} substitutions we utilize. %
Yet, as drawbacks, to make the whole solution available, we still need to deal
with the infinite terms included in $\tilde{H}(\varepsilon)$ and its
dependence on the eigenenergy $\varepsilon$ which is actually unknown before
we successfully solve the problem.

One common solution for these two problems is to employ the standard RS
perturbation method, which utilizes the expansions of $\varepsilon$  and
$|\varphi_{0}\rangle$ in Eqs. (\ref{eq:eps_exp}) and  (\ref{eq:phi_exp}), respectively,
and all possible results can be achieved by checking terms on the same
order of $\beta$ in Eqs. (\ref{eq:G_eps_def}) and (\ref{eq:eff:pseudo}).
This approach mixes up the BW and RS perturbation methods and benefits at least
on two aspects due to a fact that perturbation effects of different strengths
are able to coexist in one photon-transition matrix element which we can
manipulate in a more physical manner. %
One is that instead of the step-by-step style we directly expand Eq.
(\ref{eq:eff:pseudo}) on a specific order of $\beta^{1/4}$ and, consequently,
achieve a series of equations including all of the cases below
this order. In this context, our method now acts as an improved wrapper for the
order analysis utilized by Ref. \onlinecite{Alec2005}, and the difference is that
we use the projections before the order comparisons while they prefer that the
latter one goes first. %
The other is the convenience that we can more easily predict characteristics of
the perturbation results. For example, without the emphasis on $s\hbar\omega_{LC}$ and
the consequential result Eq. (\ref{eq:G_eps_def}), it is not obvious in the
previous paper that the projected states on
the excited levels can be derived from $|\varphi_{0}\rangle$, although the term
$s\hbar\omega_{LC}|\varphi_{s}^{(k+2)}\rangle$ in the expansion of Eq.
(\ref{eq:eigen_exp_s}) on $O(\beta^{k/4})$ with $k$ being an integer gives a hint
in the RS perturbation method. %
To avoid that $\varepsilon$ and $|\varphi\rangle$ should be obtained in pair step by step in this method,
we introduce a better one where the effective quantum states for the
system are able to share a unique set of effective operators such as the
effective Hamiltonian and the loop current operator.

\subsection{Effective Hamiltonian}
\label{sec:ph:ef}

\begin{figure}[htpb]
    \begin{center}
        \includegraphics[keepaspectratio=true,width=.5\textwidth]{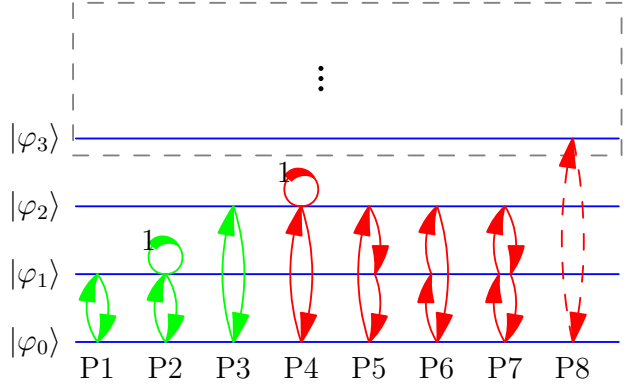}
    \end{center}
    \caption{\mycolor{}CPTPs of different types.
The last steps of those CPTPs do not denote the $\hat{\mathcal{H}}_{0,s}$-like operators but  the $\hat{{H}}_{0,s}$-like ones.}
    \label{fig:paths}
\end{figure}

\begin{table}
\caption{
For typical CPTPs labeled in Fig. \ref{fig:paths}, their
corresponding operators and the orders of their dominant terms on $\beta$ are
listed. The CPTPs P1, P3 and P8 denote the cases of the direct connection
type for the first, second and third excited levels, respectively. By checking
the orders, it is found that P1 is stronger than any CPTP involving the first
excited level such as P2, P5, P6 and P7. The photon energy denominator in the
self-transition operator ${\hat{\mathcal{H}}_{1,1}}(\varepsilon)$ yields that P2
is weaker than P1. }
\label{tab:ptps}
\begin{ruledtabular}
\begin{tabular}{lll}
Label&Operator&Order\\
\hline
P1 & $\hat{\mathcal P}_{1}=\hat{H}_{0,1}{\hat{\mathcal{H}}_{1,0}}$ & $\beta^{1}$\\
P2 & $\hat{\mathcal P}_{2}=\hat{H}_{0,1}{\hat{\mathcal{H}}_{1,1}}(\varepsilon){\hat{\mathcal{H}}_{1,0}}$ & $\beta^{3/2}$\\
P3 & $\hat{\mathcal P}_{3}=\hat{H}_{0,2}{\hat{\mathcal{H}}_{2,0}}$ & $\beta^{3/2}$\\
P4 & $\hat{\mathcal P}_{4}=\hat{H}_{0,2}{\hat{\mathcal{H}}_{2,2}}(\varepsilon){\hat{\mathcal{H}}_{2,0}}$ & $\beta^{2}$\\
P5 and P6 & $\hat{\mathcal P}_{5}=\hat{\mathcal P}_{6}^{\dag}=\hat{H}_{0,1}{\hat{\mathcal{H}}_{1,2}}{\hat{\mathcal{H}}_{2,0}}$ & $\beta^{2}$\\
P7 & $\hat{\mathcal P}_{7}=\hat{H}_{0,1}{\hat{\mathcal{H}}_{1,2}}{\hat{\mathcal{H}}_{2,1}}{\hat{\mathcal{H}}_{1,0}}$ & $\beta^{5/2}$\\
P8 & $\hat{\mathcal P}_{8}=\hat{H}_{0,3}{\hat{\mathcal{H}}_{3,0}}$ & $\beta^{2}$\\
\end{tabular}
\end{ruledtabular}
\end{table}

The photon transition path concept leads to an easier understanding on
$\tilde{H}(\varepsilon)$. Let us expand $\tilde{H}(\varepsilon)$ to order
$\beta^{3/2}$ with the aid of Fig.  \ref{fig:paths} and Table
\ref{tab:ptps}. To begin with, like $\hat{H}_{s,s_{1}}$ in Eq.
(\ref{eq:Ham_mat}), each CPTP operator in $\tilde{H}(\varepsilon)$ holds its
own identical dominant term, the order of which facilitates comparing its
relative strength with others. According to the order analysis ( see
Appendix \ref{sec:order} for some details ), all of the CPTPs involving the
third or higher excited levels, among which the one P8 $\hat{\mathcal
P}_{8}=\hat{H}_{0,3}{\hat{\mathcal{H}}_{3,0}}$ provides the maximum correction
on $O(\beta^{2})$, can be dropped as well as the infinite weak ones bound in
the three lowest levels, i.e., from P4 to P7, and a sum of the remaining
three ones P1, P2 and P3 yields one approximate pseudo-Hamiltonian
$\tilde{H}^{(3/2)}(\varepsilon)$ as
\begin{eqnarray}
  \label{eq:H_ord_6_4_1}
  \tilde{H}^{(3/2)}(\varepsilon)&=&%
  \hat{H}_{0,0}-\hat{H}_{0,1}\frac{\hat{H}_{1,0}}{\hbar\omega_{LC}}+\hat{H}_{0,1}\frac{(\hat{H}_{1,1}-\varepsilon)\hat{H}_{1,0}}{(\hbar\omega_{LC})^2}-\hat{H}_{0,2}\frac{\hat{H}_{2,0}}{2\hbar\omega_{LC}},
\end{eqnarray}
where its superscript ``$(3/2)$'' annotates that it is expanded on $O(\beta^{3/2})$
and partial higher order terms are also included.
The equation
$\tilde{H}^{(3/2)}(\varepsilon)|\varphi_{0}\rangle=\varepsilon|\varphi_{0}\rangle$
becomes
a generalized eigen-problem
\begin{eqnarray}
  \label{eq:H_ord_6_4_eq}
\tilde{H}^{(3/2)}_{L}|\varphi_{0}\rangle&=&\varepsilon\hat{R}|\varphi_{0}\rangle+O(\beta^{7/4}),
\end{eqnarray}
where
\begin{eqnarray}
  \label{eq:H_ord_6_4_eq_L}
  \tilde{H}^{(3/2)}_{L}&=& \hat{H}_{0,0}-\frac{\hat{H}_{0,1}\hat{H}_{1,0}}{\hbar\omega_{LC}}+\frac{\hat{H}_{0,1}\hat{H}_{1,1}\hat{H}_{1,0}}{(\hbar\omega_{LC})^2}-\frac{\hat{H}_{0,2}\hat{H}_{2,0}}{2\hbar\omega_{LC}},\\
  \label{eq:H_ord_6_4_eq_R}
  \hat{R}&=&1+\frac{\hat{H}_{0,1}\hat{H}_{1,0}}{(\hbar\omega_{LC})^2}.
\end{eqnarray}

Although Eq. (\ref{eq:H_ord_6_4_eq}) can be solved (see Appendix \ref{sec:sol}
for details), an alternative but more general way to eliminate the
$\varepsilon$-dependence is to substitute %
$\tilde{H}(\varepsilon)$ for $\varepsilon$ in the perturbation terms of $\tilde{H}(\varepsilon)$.
For instance, to deal with the term
$-\frac{\hat{H}_{0,1}\varepsilon\hat{H}_{1,0}}{(\hbar\omega_{LC})^2}|\varphi_0\rangle$
, we can multiply $\varepsilon$, a constant number
commuting with any operator, and $|\varphi_0\rangle$ first of all, and then replace $\varepsilon|\varphi_0\rangle$ with
$\tilde{H}(\varepsilon)|\varphi_0\rangle$ as follows,
\begin{eqnarray}
  \label{eq:replaced}
  -\frac{\hat{H}_{0,1}\varepsilon\hat{H}_{1,0}}{(\hbar\omega_{LC})^2}|\varphi_0\rangle=-\frac{\hat{H}_{0,1}\hat{H}_{1,0}}{(\hbar\omega_{LC})^2}\tilde{H}(\varepsilon)
  |\varphi_0\rangle{=}-\frac{\hat{H}_{0,1}\hat{H}_{1,0}}{(\hbar\omega_{LC})^2}\hat{H}_{0,0}|\varphi_0\rangle+{O}(\beta^{7/4}),
\end{eqnarray}
where %
only $\hat{H}_{0,0}$ in $\tilde{H}(\varepsilon)$ is kept in the final expansion. Therefore,
$\tilde{H}^{(3/2)}(\varepsilon)$ gets rid of its $\varepsilon$-dependence but
changes to a non-Hermitian effective operator $\tilde{H}^{(3/2)}_{nh}$ as
\begin{eqnarray}
  \label{eq:H_ord_6_4_2}
 \tilde{H}^{(3/2)}_{nh}&=&
\tilde{H}^{(3/2)}_{L}-\frac{\hat{H}_{0,1}\hat{H}_{1,0}\hat{H}_{0,0}}{(\hbar\omega_{LC})^2}.
\end{eqnarray}
Generally,
because this kind of substitutions can continue to increase the
orders of $\beta$ of the remaining $\varepsilon$-dependent terms
in $\tilde{H}(\varepsilon)$ until the result does not depend on $\varepsilon$ on the order we
want, this approach, namely the $\varepsilon$-$\tilde{H}(\varepsilon)$
substitution, can formally achieve an accurate and $\varepsilon$-independent operator
$\tilde{H}_{nh}$ which, however, loses its Hermiticity
completely just like $\tilde{H}^{(3/2)}_{nh}$, its expansion on $O(\beta^{3/2})$. %
As discussed in the
previous papers\cite{Klein73,Duan2001,Cherny2004}, the non-Hermiticity
comes with that $|\varphi_{0}\rangle$ is not a good effective state
candidate in the equation
\begin{eqnarray}
  \label{eq:H_nh}
  \tilde{H}_{nh}|\varphi_{0}\rangle&=&\varepsilon|\varphi_{0}\rangle.
\end{eqnarray}
If we introduce another eigenstate $|\psi\rangle$ with its eigenenergy being $\varepsilon_{\psi}$ and $\varepsilon$ is rewritten as $\varepsilon_{\varphi}$ for the sake of symmetry,
there exists an identity overlap problem as
\begin{eqnarray}
\langle\psi_{0}|\varphi_{0}\rangle\neq\delta_{\psi,\varphi},
\end{eqnarray}
which is also
indicated by the generalized eigen-problem Eq. (\ref{eq:H_ord_6_4_eq}). In
fact, defining an operator vector
$\vec{\mathcal{G}}=[\hat{\mathcal{G}}_{0},\hat{\mathcal{G}}_{1},\cdots]^{T}$
(Analogously, one can also drop the $\varepsilon$-dependence of the operators $\hat{\mathcal
G}_{1}(\varepsilon)$, $\hat{\mathcal
G}_{2}(\varepsilon)$, $\cdots$ as we do in Appendix \ref{sec:op_exp}) with its norm being
\begin{eqnarray}
\label{eq:G_n}
\hat{\mathcal{G}}_{||}=\hat{\mathcal{G}}^{\dag}_{||}=\left(\sum_{s}\hat{\mathcal{G}}_{s}^{\dag}\hat{\mathcal{G}}_{s}\right)^{\frac{1}{2}},
\end{eqnarray}
it is found that the orthogonality of the three-phase states
$\langle\psi|\varphi\rangle=\delta_{\psi,\varphi}$ in the manifold
$\mathcal{M}_{0}$ can be expressed by the components $\hat{\mathcal{G}}_{||}|\psi_0\rangle$ and $\hat{\mathcal{G}}_{||}|\varphi_0\rangle$ as
\begin{eqnarray}
\label{eq:orth_2p}
\delta_{\psi,\varphi}%
&=& \langle\psi_0|\hat{\mathcal{G}}_{||}^\dag\hat{\mathcal{G}}_{||}|\varphi_0\rangle.
\end{eqnarray}
Let us construct a new equation from Eq. (\ref{eq:H_nh}) as
\begin{eqnarray}
  \label{eq:H_nh_new}
  \tilde{H}_{eff}|\varphi_{eff}\rangle&=&\varepsilon|\varphi_{eff}\rangle
\end{eqnarray}
with the two-phase effective state $|\varphi_{eff}\rangle$ being
\begin{eqnarray}
 \label{eq:state_eff}
    |\varphi_{eff}\rangle=\hat{\mathcal{G}}_{||}|\varphi_{0}\rangle
\end{eqnarray}
and the operator $\tilde{H}_{eff}$ being
\begin{eqnarray}
\label{eq:heff}
\tilde{H}_{eff}=\hat{\mathcal{G}}_{||}\tilde{H}_{nh}\hat{\mathcal{G}}_{||}^{-1}.
\end{eqnarray}
According to Eq. (\ref{eq:orth_2p}), we have restored the
orthogonality of the effective states. Fortunately, the operator
$\tilde{H}_{eff}$ is also Hermitian (see Appendix \ref{sec:pf_herm} for the
proofs). Therefore, the effective Hamiltonian $\tilde{H}_{eff}$ with
$|\varphi_{eff}\rangle$ can describe the manifold $\mathcal{M}_{0}$ of the three-phase system \textit{accurately} in a
compact two-phase subspace.

We here give some comments on the availabilities of this method via a comparison to
the UT method. According to the above definitions, it is not difficult to
obtain
\begin{eqnarray}
  \label{eq:htr_heff}
  \hat{H}_{tr}\vec{\mathcal{G}}\hat{\mathcal{G}}_{||}^{-1}|\varphi_{eff}\rangle=\vec{\mathcal{G}}\hat{\mathcal{G}}_{||}^{-1}\tilde{H}_{eff}|\varphi_{eff}\rangle.
\end{eqnarray}
Since the state $|\varphi_{eff}\rangle$ is arbitrary and the vector
$\vec{\mathcal{G}}\hat{\mathcal{G}}_{||}^{-1}$ is explicitly unitary, our
method exactly focuses on the manifold $\mathcal{M}_{0}$
and presents a formal solution on its corresponding eigenvector belonging to
the transformation $\hat{T}$ in the UT method. So $\tilde{H}_{0}$ and $|\tilde\varphi_0\rangle$ in the UT method are equivalently $\tilde{H}_{eff}$ and $|\varphi_{eff}\rangle$ here, respectively. The UT method achieves
the expanded operator $\hat{S}$ instead of $\hat{T}=e^{i\hat{S}}$, suggesting that it may work more efficiently when the order becomes higher.
The PTP approach, however, gives clear pictures to handle
the expansions on lower orders and also successfully predicts the
properties of this problem. For instance, %
since the well-known optical selection rules forbid the photons to take %
odd times of creating and annihilating processes to go back to the same level
and since the corresponding operators $\hat{a}^{\dag}$ and $\hat{a}$ are always associated
with a factor proportional to $\beta^{1/4}$, it is found that there only
exist non-zero terms on the orders of $\beta^{1/4}$ to \textit{even} powers in $\tilde{H}_{eff}$,
some hints on which have been given by the orders of the dominant terms in the
CPTPs shown in Table \ref{tab:ptps}. See Appendix \ref{sec:order} for more details. %
So with an arbitrary integer $r$, we have $\tilde{H}_{eff}$
in the expansion to order $\beta^{r/2}$ as
\begin{eqnarray}
  \label{eq:eff_H_eq0}
  \tilde{H}_{eff}^{(r/2)}&=&\sum_{k=0}^{r}\tilde{H}_{eff}|_{\beta^{k/2}}+O(\beta^{\frac{r+1}{2}}),
\end{eqnarray}
where the operator $\tilde{H}_{eff}|_{\beta^{k/2}}$ is in proportion to
${\beta^{k/2}}$. %
In our method, the effective Hamiltonian $\tilde{H}_{eff}^{(r/2)}$ may
selectively keep some higher order terms for easier calculations and
denotations, but the
non-trivial terms $\tilde{H}_{eff}|_{\beta^{k/2}}$ for $k=0,1,\cdots,r$ are
uniquely determined and it still bears an error on $\beta^{{(r+1)}/{2}}$. %
Consequentially, one can have the eigen-problem with
improved conditions as
\begin{eqnarray}
  \label{eq:eff_eps_eq0}
  \varepsilon^{(2k+1)}\equiv0,\\
  \label{eq:eff_phi_eq0}
  |\varphi_{0}^{(2k+1)}\rangle\equiv0,
\end{eqnarray}
where $k$ is an integer, and both $\varepsilon^{(k)}$ and
$|\varphi_{0}^{(k)}\rangle$ are in proportion to $\beta^{k/4}$ defined in Eqs.
(\ref{eq:eps_exp}) and (\ref{eq:phi_exp}), respectively. %

Back to Eq. (\ref{eq:H_ord_6_4_2}), with the method provided by Eq.
(\ref{eq:H_nh_new}) and the expansion of $\hat{\mathcal{G}}_{||}$ on $O(\beta^{3/2})$ being
\begin{eqnarray}
  \label{eq:G_6_4}
  \hat{\mathcal{G}}_{||}^{(3/2)}&=&1+\frac{\hat{H}_{0,1}\hat{H}_{1,0}}{2(\hbar\omega_{LC})^2},
\end{eqnarray}
we have the effective Hamiltonian
$\tilde{H}_{eff}^{(3/2)}=\left(\tilde{H}_{eff}^{(3/2)}\right)^{\dag}$ as
\begin{eqnarray}
  \label{eq:H_eff_6_4}
  \tilde{H}^{(3/2)}_{eff}&=&
  \hat{H}_{0,0}+\tilde{H}_{0,1,0}+\tilde{H}_{0,2,0}+\tilde{H}^{(3/2)}_{3},
\end{eqnarray}
where %
\begin{eqnarray}
  \tilde{H}_{0,1,0}&=& -\frac{\hat{H}_{0,1}\hat{H}_{1,0}}{\hbar\omega_{LC}},\\
  \tilde{H}_{0,2,0}&=& -\frac{\hat{H}_{0,2}\hat{H}_{2,0}}{2\hbar\omega_{LC}},\\
  \label{eq:H_eff_6_4_3}
  \tilde{H}^{(3/2)}_{3}&=& \frac{\hat{H}_{0,1}\hat{H}_{1,1}\hat{H}_{1,0}}{(\hbar\omega_{LC})^2}-\frac{1}{2}\frac{\hat{H}_{0,1}\hat{H}_{1,0}\hat{H}_{0,0}+\hat{H}_{0,0}\hat{H}_{0,1}\hat{H}_{1,0}}{(\hbar\omega_{LC})^2}.
\end{eqnarray}

Let us scrutinize the terms in the effective Hamiltonian $\tilde{H}_{eff}^{(3/2)}$. The first term $\hat{H}_{0,0}$=$\langle\Omega_0|\hat{H}_{tr}|\Omega_0\rangle$
originates from the projection on the ground level of the oscillator.
Besides including the inductance-free two-phase Hamiltonian $\hat{H}_{0}$, it also takes into
account the vacuum fluctuations of the oscillator, both of which in total read
\begin{eqnarray}
    \label{eq:H_0_0}
    \hat{H}_{0,0}&=&\frac{1}{2}{\hat{\mathbf{Q}}_{\theta}^{T}\mathbf{C}_{2p}^{-1}\hat{\mathbf{Q}}_\theta}-\sum_{k=1}^3e^{-\gamma_k^2/2}E_{Jk}\cos\theta_k,
\end{eqnarray}
where $\gamma_k$ is a dimensionless factor as
\begin{eqnarray}\label{eq:gamma}
     \gamma_k&=&\frac{\alpha_{ser}}{\alpha_k}\sqrt[4]{\frac{2\beta}{g\alpha_{ser}}},k=1,2,3
\end{eqnarray}
with a ratio parameter
\begin{eqnarray}
  g=\frac{E_{J0}}{E_{C0}}=\frac{\Phi_{0}I_{C0}C_{0}}{\pi e^{2}}
\end{eqnarray}
showing a typical Josephson energy $E_{J0}$ compared to the charging energy $E_{C0}$. Because the sinusoidal potential of each
junction is equal to zero on average, the vacuum fluctuations equivalently flush the
junction energy $E_{Jk}$ into a weaker one $E'_{Jk}$ as
\begin{eqnarray}
    \label{eq:E_JK}
    E'_{Jk}&=&e^{-\gamma_k^2/2}E_{Jk},k=1,2,3,
\end{eqnarray}
which indicates that the effective size of the $k$th junction  is reduced by a factor
$e^{-\gamma_k^2/2}$ which gives a correction maximized on $O(\beta^{1/2})$.

The non-positive term $\tilde{H}_{0,1,0}$ providing main effects on
$O(\beta)$ relates to the interactions between the two lowest
levels of the oscillator. The states in the manifold $\mathcal{M}_{0}$ with $\varepsilon\ll\hbar\omega_{LC}$ hardly
afford one high-frequency LC photon so that the first excited level of the
oscillator is almost empty due to the ensuing energy punishment. %
Since the state occupying the
ground level can spread into the first excited level and also accept its
feedbacks {due to} the bidirectional transitions $\hat{H}_{0,1}$
and $\hat{H}_{1,0}${ between those two levels brought by the two-phase flux qubit
system}, the almost empty excited level acts as a ``mirror'' for the ground one, which endows
a correction $\tilde{H}_{0,1,0}$ to minimize the eigenenergy of the eigenstates.
According to Appendix \ref{sec:op_exp}, we have
\begin{eqnarray}
  \tilde{H}_{0,1,0}&=&
  -\frac{1}{2}L\left(\tilde{I}_{\phi}^{(2)}\right)^2,%
\end{eqnarray}
where the $\beta$-independent current operator
\begin{eqnarray}
  \tilde{I}_{\phi}^{(0)}&=&
  C_{ser}\sum_{k=1}^3\frac{I_{Ck}}{C_k}\sin\theta_k\label{eq:I_phi_0}
\end{eqnarray}
resembles $I_{loop}(t)$ in Eq. (\ref{eq:AC_time}) with $\phi_{1}$,
$\phi_{2}$ and $\phi_{3}$ being replaced by the effective phase variables
$\theta_{1}$, $\theta_{2}$ and $\theta_{3}$, respectively, and dominates in
\begin{eqnarray}
  \tilde{I}_{\phi}^{(2)}&=&
  C_{ser}\sum_{k=1}^3\frac{e^{-\gamma_k^2/2}I_{Ck}}{C_k}\sin\theta_k,\label{eq:I_phi_2}
\end{eqnarray}
which equivalently keeps the critical current of the $k$th junction
modified by a fluctuation factor $e^{-\gamma_k^2/2}$ like the case of $E'_{Jk}$. It is worth noting that the coupling
$\hat{V}_{1}\hat{\phi}=\frac{\Phi_0\hat{\phi}}{2\pi}\tilde{I}_{\phi}^{(0)}$
in $\hat{H}_{int}$ interestingly renders $\tilde{I}_{\phi}^{(0)}$ and also
presents the dominant terms in $\hat{H}_{0,1}$ and $\hat{H}_{1,0}$. %
To emphasize it, we assume that those two subsystems couple with each other only by
$\hat{V}_{1}\hat{\phi}$ and have the Hamiltonian
\begin{eqnarray}
  \hat{H}_{tr,\phi}&=&\hat{H}_{0}-\frac{1}{2}L\left(\tilde{I}_{\phi}^{(0)}\right)^2+\hat{D}_{\hat{a}^\dag\hat{a}},
\end{eqnarray}
where the operator
\begin{eqnarray}
  \label{eq:Daa}
  \hat{D}_{\hat{a}^\dag\hat{a}}&=& \frac{\hat{Q}^2_\phi}{2C_{ser}}+\frac{1}{2L}\left(\frac{\Phi_{0}}{2\pi}\hat\phi+L\tilde{I}_{\phi}^{(0)}\right)^{2}
\end{eqnarray}
indicates an LC-oscillator with an additional flux displacement
$-L\tilde{I}_{\phi}^{(0)}$. %
In a semi-classical picture, the above formula suggests that
the average value of the current in the loop inductance is expected to be
$\tilde{I}_{\phi}^{(0)}$ as a function of the slow junction phases
instead of a real zero value when $L\rightarrow0$, so $\tilde{I}_{\phi}^{(0)}$
can be understood as the loop current produced by the junctions which drives
the inductance to generate an additional small flux. As a result of that the
slow-varying-function biased LC-oscillator does not change its own
eigenenergy significantly, the inductive energy
$-\frac{1}{2}L\left(\tilde{I}_{\phi}^{(0)}\right)^2$ on $O(\beta)$ is added as
one perturbation correction to the effective two-phase Hamiltonian, which can be
explained as that the flux generated by $\tilde{I}_{\phi}^{(0)}$ in the inductance also
affects the junctions themselves. Intuitively, this self-bias effect persistently
lowers the potential on any point which keeps a non-zero current and always
opposes the current direction switching. In the
quantum regime, this kind of understanding is still supported by the facts that
the estimation
$\langle\varphi|\hat{D}_{\hat{a}^\dag\hat{a}}|\varphi\rangle=\frac{1}{2}\hbar\omega_{LC}+O(\beta^{3/2})$
provides no effect on $O(\beta)$ and that the inductive energy correction dominates
in $\tilde{H}_{0,1,0}$.  Furthermore, a rigorous analysis
in the next section also confirms that $\tilde{I}_{\phi}^{(0)}$ is the loop
current operator for the inductance-free flux qubit.

The term $\tilde{H}_{0,2,0}$ shows the
direct interactions between the ground and the second excited levels of the
LC-oscillator via two-photon transitions. Photons travel forth and back
via the bidirectional transitions $\hat{H}_{0,2}$ and $\hat{H}_{2,0}$,
resulting in a non-positive operator
\begin{eqnarray}
  \label{eq:h02h20}
  \frac{\tilde{H}_{0,2,0}}{E_{J0}}&=& -\frac{1}{4}\sqrt{\frac{\alpha_{ser}^{7}\beta^{3}}{2g}}\left(\sum_{k=1}^{3}\frac{e^{-\frac{\gamma^2_k}{2}}}{\alpha_k}\cos\theta_{k}\right)^2,
\end{eqnarray}
according to Appendix \ref{sec:op_exp}. Its main effects are on
$O(\beta^{3/2})$ contributed by the coupling $\hat{V}_{2}\hat\phi^{2}\neq0$ in
$\hat{H}_{int}$.

Finally, the last term %
$\tilde{H}^{(3/2)}_{3}$ corresponds to the CPTP operator $\hat{\mathcal
P}_{2}=\hat{H}_{0,1}{\hat{\mathcal{H}}_{1,1}}(\varepsilon){\hat{\mathcal{H}}_{1,0}}$
which includes the self-transition ${\hat{\mathcal{H}}_{1,1}}(\varepsilon)$ of
the first excited level.
Since the capacitive energy part
$\frac{1}{2}{\hat{\mathbf{Q}}_{\theta}^{T}\mathbf{C}_{2p}^{-1}\hat{\mathbf{Q}}_\theta}$
of the unperturbed Hamiltonian $\hat{H}_{0}$ does not commute with
$\hat{H}_{1,0}$ and $\hat{H}_{0,1}$ which turn out as functions of the
effective phase variables $\hat{\theta}_{1}$ and $\hat{\theta}_{2}$, in the
effective phase representation ($\theta_{1}$,$\theta_{2}$) with
\begin{eqnarray}
  \label{eq:2fold_commu}
  \left[ \left[ \frac{\partial}{\partial\theta_{s}}\frac{\partial}{\partial\theta_{t}},f  \right],f\right]&=& 2\frac{\partial{f}}{\partial\theta_{s}}\frac{\partial{f}}{\partial\theta_{t}},\text{for }s,t=1,2,
\end{eqnarray}
simplifying the right hand side of  Eq.  (\ref{eq:H_eff_6_4_3}) yields $\tilde{H}^{(3/2)}_{3}$ in a symmetric form for the three junctions as
\begin{eqnarray}
  \label{eq:H_eff_6_4_3_I}
  \tilde{H}^{(3/2)}_{3}&=& -\frac{L}{8\hbar\omega_{LC}}\left[\left[{\hat{\mathbf{Q}}_{\theta}^{T}\mathbf{C}_{2p}^{-1}\hat{\mathbf{Q}}_\theta},\tilde{I}_{\phi}^{(0)}\right],\tilde{I}_{\phi}^{(0)}\right]+O(\beta^{2}),\\
  \label{eq:H_eff_6_4_3_3}
  \tilde{H}^{(3/2)}_{3}/f&=&
  \sum^{3}_{k=1}\alpha_{k}\sum^{3}_{k=1}\cos^2\theta_{k}-\sum^{3}_{k=1}\alpha_{k}\cos^2\theta_{k}-2\prod^{3}_{k=1}\cos\theta_{k}\sum^{3}_{k=1}\frac{\alpha_{k}}{\cos\theta_{k}}+O(\beta^{2}),
\end{eqnarray}
where
\begin{eqnarray}
  \label{eq:H_eff_6_4_3_4}
  f&=& {\sqrt{\frac{\alpha_{ser}^{7}\beta^{3}}{2g\prod^{3}_{k=1}\alpha_{k}^{2}}}}E_{J0}.%
\end{eqnarray}

Therefore, the effective Hamiltonian $\tilde{H}^{(3/2)}_{eff}$ has taken
account of four corrections of different types to the unperturbed
one $\hat{H}_{0}$. Its complicated expression indicates that treating
the LC-oscillator as a three-level system does not stand as an easy task on the derivations
and analysis. %
First of all, unlike common perturbation situations where two subsystems
couple with each other via a weak linear interaction, the Josephson
junctions exhibiting as nonlinear inductances keep the %
interaction $\hat{H}_{int}$ in Eq.(\ref{eq:H_int}) %
split into the couplings of different strengths. For instance, among the effective corrections
in proportion to $\beta^{3/2}$, $\hat{V}_{6}\hat\phi^6$ donates one as
$\hat{V}_{6}\langle\Omega_0|\hat\phi^6|\Omega_0\rangle$ in $\hat{H}_{0,0}$, and
$\hat{V}_{3}\hat\phi^3$ as $-{(\hbar\omega_{LC})}^{-1}{\hat{V}_{3}\langle\Omega_0|\hat\phi^3|\Omega_1\rangle\hat{V}_{1}\langle\Omega_1|\hat\phi|\Omega_0\rangle}$
in $\tilde{H}_{0,1,0}$.
This kind of terms inside the photon-transition operators $\hat{H}_{s,s_{1}}$ have been automatically
included in our results while the step-by-step method should explicitly
calculate them out. On the other hand, $\hat{V}_{2}\hat\phi^2$ turns on the
direct connections between the second excited level and the ground one, thus
straightforwardly imposing the influences of this excited level without the help of any other excited one; otherwise, only with
$\hat{V}_{1}\hat\phi$ its maximum feedback decreases to the CPTP P7
$\hat{\mathcal
P}_{7}=\hat{H}_{0,1}{\hat{\mathcal{H}}_{1,2}}{\hat{\mathcal{H}}_{2,1}}{\hat{\mathcal{H}}_{1,0}}$
in Table \ref{tab:ptps}, the dominant term of which is on $O(\beta^{5/2})$. Thus
our method also needs to accumulate suitable CPTPs one by one. %
Moreover, although being a part of the effective potential in
$\tilde{H}_{eff}^{(3/2)}$, the operator $\tilde{H}^{(3/2)}_{3}$ involves one
self-transition process and performs as a correction sensitive to the eigenenergy (
see the pseudo-Hamiltonian $\tilde{H}^{(3/2)}(\varepsilon)$ ). %
This feature is not good for the analysis of the experiments which
often alter the energy level structure of the whole system by changing the
external flux bias. %
After solving the eigen-problem of $\tilde{H}^{(3/2)}_{eff}$, although the eigenvalue
$\tilde\varepsilon^{(3/2)}$ directly gives
\begin{eqnarray}
  \label{eq:eff_eps_3_2}
  \varepsilon=\tilde\varepsilon^{(3/2)}+O(\beta^{2}),
\end{eqnarray}
the effective state $|\varphi_{eff}^{(3/2)}\rangle$ should be preprocessed
as %
\begin{eqnarray}
  \label{eq:phi0_eq_eff}
  |\varphi_{0}\rangle&=& \left(1-\frac{\hat{H}_{0,1}\hat{H}_{1,0}}{2(\hbar\omega_{LC})^2}\right)|\varphi_{eff}^{(3/2)}\rangle+O(\beta^{2})
\end{eqnarray}
for further discussions. Even if the difficulties mentioned above are carefully
handled, %
we should still cope with tens of terms related to $\alpha_{1,2,3}$, $\beta$ and $g$.
Limited by the fabrication conditions and other factors, the loop inductance cannot be enlarged too much, and thus the $O(\beta^{3/2})$-effects appear essential in rare cases.
Therefore, as a compromise between simplicity and accuracy, we choose one effective Hamiltonian on
$O(\beta)$ rougher but \textit{optimal} in this trade-off as
\begin{eqnarray}
  \label{eq:H_ord_4_4}
  \tilde{H}^{(1)}_{eff}&=&\hat{H}_{0,0}-\frac{\hat{H}_{0,1}\hat{H}_{1,0}}{\hbar\omega_{LC}},
\end{eqnarray}
which %
bears an error on $O(\beta^{3/2})$. %
Dropping the terms in proportion to $\beta^{3/2}$ or higher orders of $\beta$ in
Eq. (\ref{eq:H_ord_4_4}) yields an effective potential
\begin{eqnarray}
  \label{eq:H_ord_4_4_exp}
  \tilde{V}_{eff}^{(1)}&=& -\sum_{k=1}^3\left(1-\sqrt{\frac{\alpha_{ser}^3}{2g\alpha_k^4}}\beta^{\frac{1}{2}}+{\frac{\alpha_{ser}^3}{4g\alpha_k^4}}\beta\right)E_{Jk}\cos\theta_k-\frac{1}{2}L\left(\tilde{I}_{\phi}^{(0)}\right)^2
\end{eqnarray}
identical to the one presented by Ref.\onlinecite{Alec2005}. The
corresponding normalized effective eigenstate $|\varphi_{eff}^{(1)}\rangle$ %
approximates $|\varphi_{0}\rangle$ on $O(\beta)$ as
\begin{eqnarray}
  \label{eq:phi0_eq_eff_1}
  |\varphi_{0}\rangle&=& |\varphi_{eff}^{(1)}\rangle+O(\beta^{3/2})
\end{eqnarray}
with the eigenenergy $\tilde\varepsilon^{(1)}$ being
\begin{eqnarray}
  \label{eq:eff_eps_1}
  \varepsilon=\tilde\varepsilon^{(1)}+O(\beta^{3/2}).
\end{eqnarray}

\subsection{Arbitrary effective operator}
\label{sec:ph:ar}

As mentioned above, the photon transition path method presents not only
an accurate prediction on the eigenenergy by the effective
Hamiltonian but also a full description of how an effective
two-phase system is mapped to the three-phase one. Take an arbitrary
three-phase operator $\hat{F}$ as an example. Assume that two arbitrary
eigenstates $|\varphi\rangle$ and $|\psi\rangle$ in the manifold
$\mathcal{M}_{0}$ go with their eigenenergy $\varepsilon_{\varphi}$ and
$\varepsilon_{\psi}$, respectively, where $|\varphi\rangle$ may equate to
$|\psi\rangle$. The expansions
\begin{eqnarray}
  \label{eq:op_expn}
  \hat{F}&=& \sum_{m,n}|\Omega_m\rangle\hat{F}_{m,n}\langle\Omega_n|,\\
  |\varphi\rangle&=&\sum_{s}\left(\hat{\mathcal
  G}_{s}|\varphi_0\rangle\right)|\Omega_s\rangle,\\
  |\psi\rangle&=&\sum_{s}\left(\hat{\mathcal
  G}_{s}|\psi_0\rangle\right)|\Omega_s\rangle.
\end{eqnarray}
yield
\begin{eqnarray}
  \label{eq:op_mat}
\langle\psi|\hat{F}|\varphi\rangle&=& \langle\psi_0|\left(\sum_{m,n}\hat{\mathcal G}_{m}^\dag\hat{F}_{m,n}\hat{\mathcal G}_{n}\right)|\varphi_0\rangle\\\nonumber
&=& \langle\psi_{eff}|\tilde{F}_{eff}|\varphi_{eff}\rangle,
\end{eqnarray}
where we define the effective operator for $\hat{F}$ as
\begin{eqnarray}
  \tilde{F}_{eff}&=& \sum_{m,n}\left(\hat{\mathcal G}_{m}\hat{\mathcal{G}}_{||}^{-1}\right)^\dag\hat{F}_{m,n}\hat{\mathcal G}_{n}\hat{\mathcal{G}}_{||}^{-1}.
\label{eq:op_eff}
\end{eqnarray}
The two-phase effective operator $\tilde{F}_{eff}$ depends not only on
$\hat{F}_{m,n}$,  for example, which may obey the optical selection rules, but
also on $\hat{\mathcal G}_{m}$, $\hat{\mathcal
G}_{n}$ and $\hat{\mathcal{G}}_{||}^{-1}$ which portray all of the projected components of the eigenstates in the three-phase system. Especially, for $\hat{F}$ as the three-phase identity operator
$\hat{I}$, one can have
\begin{eqnarray}
    \tilde{I}_{eff}&=& \sum_{n}\left(\hat{\mathcal G}_{n}\hat{\mathcal{G}}_{||}^{-1}\right)^\dag\hat{\mathcal G}_{n}\hat{\mathcal{G}}_{||}^{-1}=\hat{I}_{2p},
\label{eq:I_eff}
\end{eqnarray}
where $\hat{I}_{2p}$ is the identity operator for the effective two-phase
subspace; for $\hat{F}=\hat{H}_{tr}$, with Eq. (\ref{eq:htr_heff}) we have
a self-consistent result as
\begin{eqnarray}
  \left(\hat{H}_{tr}\right)_{eff}&=&
  \left(\vec{\mathcal{G}}\hat{\mathcal{G}}_{||}^{-1}\right)^{H}\hat{H}_{tr}\vec{\mathcal{G}}\hat{\mathcal{G}}_{||}^{-1}=\tilde{H}_{eff}.
\end{eqnarray}

\section{effective current operator}
\label{sec:ef}

In the three-phase system, the current operator can be achieved in different
ways: on the one hand, the definition of the loop inductance $L$ yields that
\begin{equation}
  \label{eq:I_phi}
  \hat I_{\phi}=-\frac{\hat\Phi_{L}}{L}=-\frac{\Phi_{0}}{2\pi
  L}\hat\phi;
\end{equation}
on the other hand, according to Kirchhoff's current law, the
series current flowing towards the $k$th junction(for $k=1$, 2 and 3) can be expressed as a sum of its Josephson
supercurrent $I_{Ck}\sin\hat\phi_{k}$ and the one through the capacitor $C_k$, which due to
the time variation of the charge $\hat{Q}_{k}$ is provided by the Heisenberg equation
\begin{eqnarray}
  \label{eq:Q_k_I_phi}
  \dot{Q}_{k}&=&\left[\hat
    H_{tr}, i\hat Q_{k}\slash\hbar\right]=\hat I_{\phi}-I_{Ck}\sin\hat\phi_{k}.
\end{eqnarray}
 One can simply find that the series current
operator of the loop possesses a unique form $\hat I_{\phi}$.

The virtual work principle, besides the direct derivation above,  also suggests
some reasonable forms as
\begin{eqnarray}
\label{eq:I_par_phi}
\hat{I}_{\partial\Phi_X}&=&\frac{\partial\hat{V}_{tr}}{\partial\Phi_X},
\end{eqnarray}
where $\hat{V}_{tr}$ is the potential term of $\hat{H}_{tr}$,
because for the eigenstate $|\varphi\rangle$ we have
\begin{eqnarray}
\label{eq:I_par_eng}
\langle\varphi|\hat{I}_{\partial\Phi_X}|\varphi\rangle=\langle\varphi|\frac{\partial\hat{H}_{tr}}{\partial\Phi_X}|\varphi\rangle
=\frac{\partial{\varepsilon}_\varphi}{\partial\Phi_X}.
\end{eqnarray}
However, since the translations such as the one in
Eq. (\ref{eq:theta_trans}) can alter the dependence of $\hat{V}_{tr}$ on $\Phi_X$, lots of current operators such as $\hat{I}_{\phi}$,
the DC current operators $I_{Ck}\sin\hat\phi_{k}$ for k=1,2 and 3, etc, are
possible candidates in this approach, which all provide the same diagonal
matrix elements $\langle\varphi|\hat{I}_{\partial\Phi_X}|\varphi\rangle$ equal to $\partial{\varepsilon_{\varphi}}/\partial{\Phi_X}$.
Unfortunately, this method cannot inform us which one is proper for the
non-diagonal elements.

With the current operator being ready for the three-phase system, one can expand it
in the oscillator-subsystem as
\begin{eqnarray}
    \label{eq:I_phi_mat}
    \langle\Omega_m|\hat{I}_{\phi}|\Omega_n\rangle&=&-\sqrt[4]{\frac{\hbar^{2}}{4C_{ser}L^{3}}}\left(\sqrt{m}\delta_{m,n+1}+\sqrt{n}\delta_{m+1,n}\right).
\end{eqnarray}
With the effective theory shown in Eqs. (\ref{eq:op_mat}) and (\ref{eq:op_eff}), it yields that
\begin{eqnarray}
  \label{eq:I_phi_gen}
  \langle\psi|\hat{I}_\phi|\varphi\rangle&=&-\sqrt[4]{\frac{\hbar^{2}}{4C_{ser}L^{3}}}\sum_{k\ge1}\sqrt{k}\left(\langle\psi_{k}|\varphi_{k-1}\rangle+\langle\psi_{k-1}|\varphi_{k}\rangle\right)
\end{eqnarray}
and
\begin{eqnarray}
  \label{eq:I_eff_gen}
  \tilde{I}_\phi&=&-\sqrt[4]{\frac{\hbar^{2}}{4C_{ser}L^{3}}}\sum_{k\ge1}\sqrt{k}\left(\hat{\mathcal{G}}_{||}^{-1}\right)^\dag\left(\hat{\mathcal G}_{k}^\dag\hat{\mathcal G}_{k-1}+\hat{\mathcal G}_{k-1}^\dag\hat{\mathcal G}_{k}\right)\hat{\mathcal{G}}_{||}^{-1},
\end{eqnarray}
where the tilde symbol labels the effective operators.

Let us check the dependence of
$\tilde{I}_\phi$ on the reduced inductance $\beta$. The dimensional factor
$\sqrt[4]{\frac{\hbar^{2}}{4C_{ser}L^{3}}}$, belonging to $\hat{I}_{\phi}$ in
proportion to $\beta^{-3/4}$ , indicates that this current operator $\hat{I}_{\phi}$ generally
diverges with the loop size. For example, the state $|\varphi\rangle$ with $|\langle\varphi_{0}|\varphi_{1}\rangle|=\frac{1}{2}$ carries
an infinite loop-current $\langle\varphi|\hat{I}_{\phi}|\varphi\rangle$ when
$\beta\rightarrow0$. Oppositely, the
optical selection rules zero out any rule-breaking term regardless of the order of its scale factor on $\beta$, so a real
dark state $|\varphi_{dark}\rangle$ in the manifold $\mathcal{M}_{0}^{(0)}$ where the excited levels of the LC-oscillator are entirely empty
is forbidden to possess a current circulating in the
loop due to $\langle\Omega_{0}|\hat\phi|\Omega_{0}\rangle=0$. %
As a result, with the dimensional factor the largest term $\frac{\hat{H}_{1,0}}{\hbar\omega_{LC}}$ among the small perturbations left in the sum in Eq. (\ref{eq:I_eff_gen}) provides an inductance-independent operator $\tilde{I}_{\phi}^{(0)}$, which has been written in Eq.(\ref{eq:I_phi_0}) and confirms that the
junctions determine the loop current when the inductance is small enough. %
One interesting thing is that the effective counterpart for the photon
number operator $\hat{a}^\dag\hat{a}$,
\begin{eqnarray}
  (\hat{a}^\dag\hat{a})_{eff}&=&
  \frac{L\left(\tilde{I}_{\phi}^{(0)}\right)^2}{2\hbar\omega_{LC}}+O(\beta^{2}),
\label{eq:aa_eff}
\end{eqnarray}
is mainly determined by the inductive energy $L\left(\tilde{I}_{\phi}^{(0)}\right)^2/2$ divided by one LC-photon energy $\hbar\omega_{LC}$. %
Thus the eigenstates in the manifold $\mathcal{M}_{0}$
actually look \textit{dim} with the average photon number being much less than
one, neither dark with no photon completely nor bright with one or more
photons.

Expanding Eq. (\ref{eq:I_eff_gen}) to order $\beta$, we obtain the effective current operator $\tilde{I}_{\phi}^{(4)}$ as
\begin{eqnarray}
  \label{eq:eff_I_4_4}
  \tilde{I}_{\phi}^{(4)}&=&-\sqrt[4]{\frac{\hbar^{2}}{4C_{ser}L^{3}}}
  \left(\hat{\mathcal G}_{1}+\sqrt{2}\hat{\mathcal
  G}_{2}^{\dag}\hat{\mathcal
  G}_{1}+h.c.\right)_{O(\beta^{{7}/{4}})},
\end{eqnarray}
where the LC-oscillator is involved as a three-level system and the operators in
the parentheses such as $\hat{\mathcal G}_{1}$ should be expanded on
${O(\beta^{{7}/{4}})}$ denoted by the subscript. %
It is clear that the large dimensional factor leads to deeper explorations on
the $\hat{\mathcal G}$-operators: $\hat{\mathcal G}_{1}$ should be expanded at least on
${O(\beta^{{7}/{4}})}$ in $\tilde{I}_{\phi}^{(4)}$ but
on ${O(\beta^{{3}/{4}})}$ in  $\tilde{H}_{eff}^{(1)}$ and on ${O(\beta^{{5}/{4}})}$ in  $\tilde{H}_{eff}^{(3/2)}$. %
According to the optical selection rules, it is
found that $\tilde{I}_{\phi}^{(4)}$ suffers from an error on $O(\beta^{3/2})$
compared to $\tilde{I}_{\phi}$ as
\begin{eqnarray}
  \label{eq:eff_I_4_4_err}
  \tilde{I}_{\phi}&=&\tilde{I}_{\phi}^{(4)}+{O(\beta^{3/2})}.
\end{eqnarray}

Although the direct expansion in Eq. (\ref{eq:eff_I_4_4}) can be accomplished
with the help of Appendix \ref{sec:op_exp}, %
since our effective theory can cope with arbitrary three-phase operators, %
one can also achieve $\tilde{I}_{\phi}^{(4)}$ via applying the theory to its
another definition on $O(\beta)$. %
For the sake of clarity, $I_{C0}$ is utilized as the current
unit, $\hbar=1$ and $e=1/2$. %
We construct a current operator
\begin{eqnarray}
  \label{eq:I_3p_sigma}
  \hat{I}_{con}&=& \alpha_{ser}\sum_{k=1}^{3}\sin\hat\phi_{k}\\\nonumber
  &=& \hat{I}_{\cos}-\hat{\mathcal\eta}_{\beta}\hat{I}_\phi+\hat{I}_{\sin},
\end{eqnarray}
where according to Eqs. (\ref{eq:theta_for_short}) and (\ref{eq:theta3}) it has
been divided into three components as
\begin{eqnarray}
  \hat{I}_{\cos}&=& \alpha_{ser}\sum_{k=1}^3\sin\hat{\theta}_k\cos\frac{\alpha_{ser}}{\alpha_k}\hat{\phi},%
  \label{eq:I_cos_}%
  \\
  \hat{\mathcal\eta}_{\beta}&=& \alpha_{ser}^2\beta\sum_{k=1}^3\frac{\cos\hat{\theta}_k}{\alpha_k},%
  \label{eq:eta_beta}%
  \\
  \hat{I}_{\sin}&=& \alpha_{ser}\sum_{k=1}^3\cos\hat{\theta}_k\sin\frac{\alpha_{ser}}{\alpha_k}\hat{\phi}+\hat{\mathcal\eta}_{\beta}\hat{I}_\phi
  \label{eq:Isin}\\\nonumber
  &=& \alpha_{ser}^4\beta^3\hat{I}_\phi^3\sum_{k=1}^3\frac{\cos\hat{\theta}_k}{6\alpha_k^3}+\cdots.
\end{eqnarray}
With the aid of Eq. (\ref{eq:Q_k_I_phi}), it is not difficult to achieve that
\begin{eqnarray}
  \label{eq:H_Q_sigma_def}
  \dot{Q}_{\phi}=\left[\hat{H}_{tr},i\hat{Q}_\phi/\hbar\right]=\hat{I}_\phi-\hat{I}_{con},
\end{eqnarray}
which reads
\begin{eqnarray}
  \label{eq:dQ_III}
  \dot{Q}_{\phi}&=& \left(1+\hat{\mathcal\eta}_{\beta}\right)\hat{I}_\phi-\hat{I}_{\cos}-\hat{I}_{\sin}.
\end{eqnarray}
Applying $\left(1-\hat{\mathcal\eta}_{\beta}\right)\cdot$ to both sides of the above equation, we have
\begin{eqnarray}
  \label{eq:I_I_dQ}
  \hat{I}_\phi&=& \left(1-\hat{\mathcal\eta}_{\beta}\right)\left(\hat{I}_{\cos}+\hat{I}_{\sin}+\dot{Q}_{\phi}\right)+\hat{\mathcal\eta}_{\beta}^2\hat{I}_\phi.
\end{eqnarray}
Utilizing the definition in Eq. (\ref{eq:op_eff}),
one can find that the effective operators of the three-phase ones
$\left(1-\hat{\mathcal\eta}_{\beta}\right)\hat{I}_{\sin}$,
$\hat{\mathcal\eta}_{\beta}\dot{Q}_{\phi}=\hat{\mathcal\eta}_{\beta}\left(\hat{I}_\phi-\hat{I}_{con}\right)$ and
$\hat{\mathcal\eta}_{\beta}^2\hat{I}_\phi$ equate to zero on $O(\beta)$, and,
thus, it follows that
\begin{eqnarray}
  \tilde{I}_{\phi}^{(4)}=\tilde{I}_{\phi}^{(2)}-\hat{\mathcal\eta}_{\beta}\tilde{I}_{\phi}^{(2)}+\dot{Q}_{\phi}^{eff}+O(\beta^{3/2}),
\end{eqnarray}
where
\begin{eqnarray}
  \tilde{I}_{\phi}^{(2)}&=& \langle\Omega_0|\hat{I}_{\cos}|\Omega_0\rangle=\langle\Omega_0|\hat{I}_{con}|\Omega_0\rangle,\\
  \dot{Q}_{\phi}^{eff}&=& \frac{\left[\hat{H}_{0},\left[\hat{H}_{0},\tilde{I}_{\phi}^{(2)}\right]\right]}{\left(\hbar\omega_{LC}\right)^2}+O(\beta^{3/2}).
  \label{eq:dQ_eff}
\end{eqnarray}
Three kinds of effects are taken into account in the above formula for $\tilde{I}_{\phi}^{(4)}$. The first term
$\tilde{I}_{\phi}^{(2)}$ consistent with its definition in Eq.
(\ref{eq:I_phi_2}) shows that the projections
$\langle\Omega_0|\cos\left({\alpha_{ser}}{\alpha_k}^{-1}\hat{\phi}\right)|\Omega_0\rangle$
impose the vacuum fluctuation factors $e^{-\gamma_k^2/2}$ to the corresponding
terms in $\tilde{I}_{\phi}^{(0)}$ for $k=1$,2 and 3. %
The second term $-\hat{\mathcal\eta}_{\beta}\tilde{I}_{\phi}^{(2)}$ is traced
back to $-\hat{\mathcal\eta}_{\beta}\hat{I}_\phi$ in $\hat{I}_{con}$
as the result of the linear approximations
${\alpha_k}^{-1}{\alpha_{ser}}\hat{\phi}$ for
$\sin\left({\alpha_k}^{-1}{\alpha_{ser}}\hat{\phi}\right)$. %
And the final term $\dot{Q}_{\phi}^{eff}$ represents a tiny current
$\dot{Q}_{\phi}$ flowing through the series capacitance $C_{ser}$, which can also be
obtained from the direct expansion in Eq. (\ref{eq:eff_I_4_4}) or the solution
presented by Appendix \ref{sec:de}. As a two-fold commutator, it correspondingly
involves a self-transition process $\hat{\mathcal
H}_{1,1}^2(\varepsilon)\hat{\mathcal H}_{1,0}$ occurring in the first excited
level of the oscillator and, thus, explains why the self-transition processes
are able to challenge the Hermiticity of the effective
Hamiltonian. %
With the effective states $|\psi_{eff}^{(1)}\rangle$ and
$|\varphi_{eff}^{(1)}\rangle$, the matrix element
$\langle\psi_{eff}^{(1)}|\dot{Q}_{\phi}^{eff}|\varphi_{eff}^{(1)}\rangle$ can
be calculated numerically as
\begin{eqnarray}
  \langle\psi_{eff}^{(1)}|\dot{Q}_{\phi}^{eff}|\varphi_{eff}^{(1)}\rangle=\left(\frac{\tilde\varepsilon^{(1)}_{\psi}-\tilde\varepsilon^{(1)}_{\varphi}}{\hbar\omega_{LC}}\right)^2\langle\psi_{eff}^{(1)}|\tilde{I}_{\phi}^{(2)}|\varphi_{eff}^{(1)}\rangle+O(\beta^{3/2}),
\end{eqnarray}
where $\tilde\varepsilon^{(1)}_{\psi}$ and $\tilde\varepsilon^{(1)}_{\varphi}$
being involved indicates that the self-transition effects distinguish the
corresponding eigenstates by their different eigenenergy.

Now we have a short summary of several effective current operators in the
effective theory. The operator $\tilde{I}_{\phi}^{(0)}$ defined in Eq.
(\ref{eq:I_phi_0}) as an effective current operator excluding any inductive
effect acts as the loop current operator for the inductance-free flux qubit
system. The operator $\tilde{I}_{\phi}^{(2)}$ defined in Eq. (\ref{eq:I_phi_2})
appearing in the effective Hamiltonian $\tilde{H}_{eff}^{(1)}$ contains the vacuum fluctuation corrections while both $\tilde{I}_{\phi}^{(0)}$ and $\tilde{I}_{\phi}^{(2)}$ treat the oscillator as a
two-level system. The third one $\tilde{I}_{\phi}^{(4)}$, costing
more, can include the effects brought by the second excited level of the oscillator and, especially, possesses a term
coming from a self-transition process which the effective Hamiltonians
$\tilde{H}_{eff}^{(1)}$ and $\tilde{H}_{eff}^{(3/2)}$ do not have. Formally, the effective current operator $\tilde{I}_{\phi}$ like $\tilde{H}_{eff}$
can be truncated on arbitrary orders. However, for the next step, to achieve the
fourth one $\tilde{I}_{\phi}^{(6)}$ accurate on $O(\beta^{3/2})$, we should no
longer neglect $\hat{\mathcal{G}}_{||}^{-1}$ and expanding
$\hat{\mathcal{G}}_{1}$ to order $\beta^{9/4}$ turns out as a more cumbersome task
without any surprise. In the trade-off between simplicity and accuracy,
we choose $\tilde{I}_{\phi}^{(4)}$ as the \textit{optimal} approximation
for $\tilde{I}_{\phi}$, which is also justified by the
following numerical simulations.

\section{numerical discussion}
\label{sec:br}
\begin{figure}
\begin{minipage}[c]{0.3\textwidth}
\centering
\includegraphics[keepaspectratio=true,width=.95\textwidth]{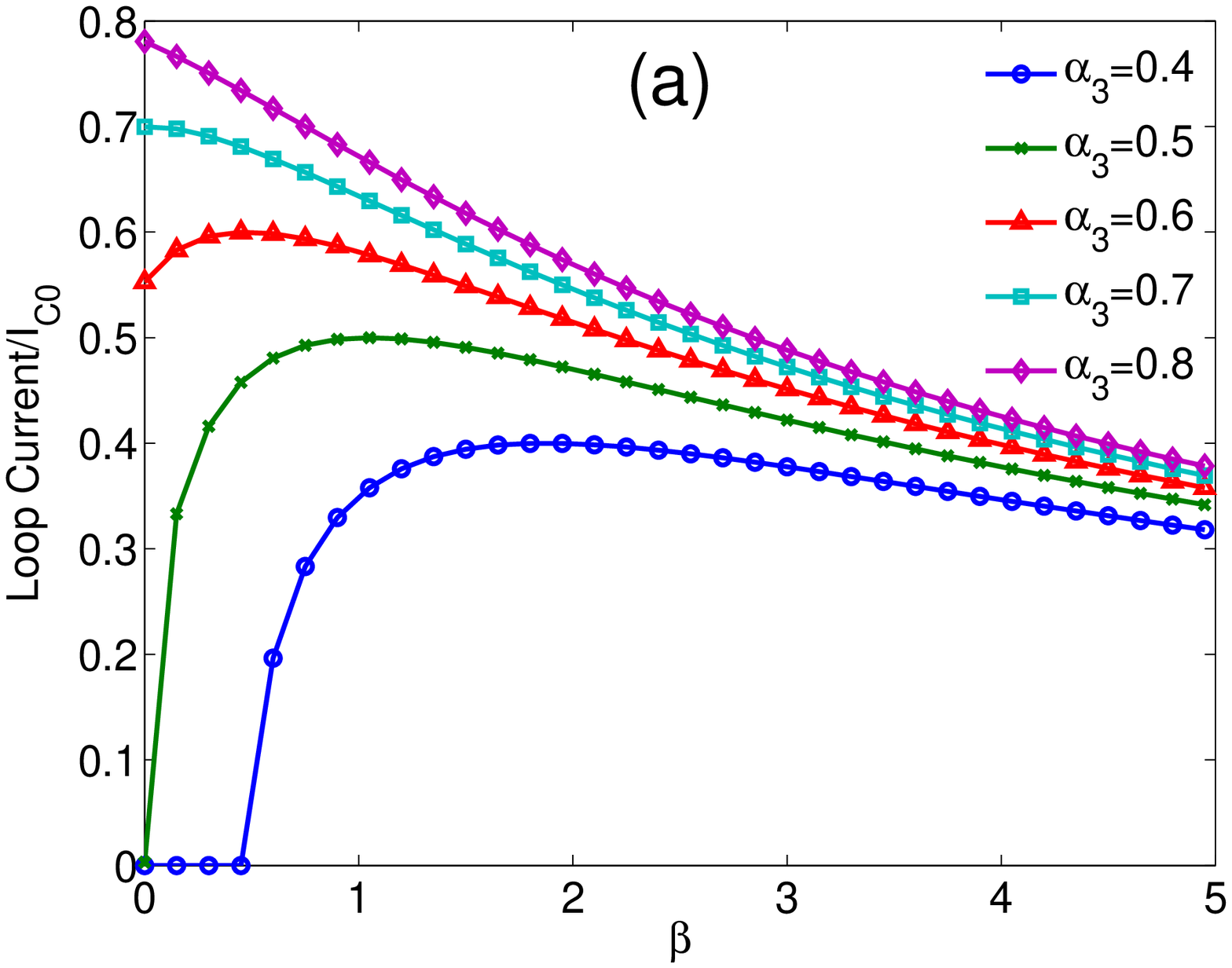}
\end{minipage}%
\begin{minipage}[c]{0.3\textwidth}
\centering
\includegraphics[keepaspectratio=true,width=.95\textwidth]{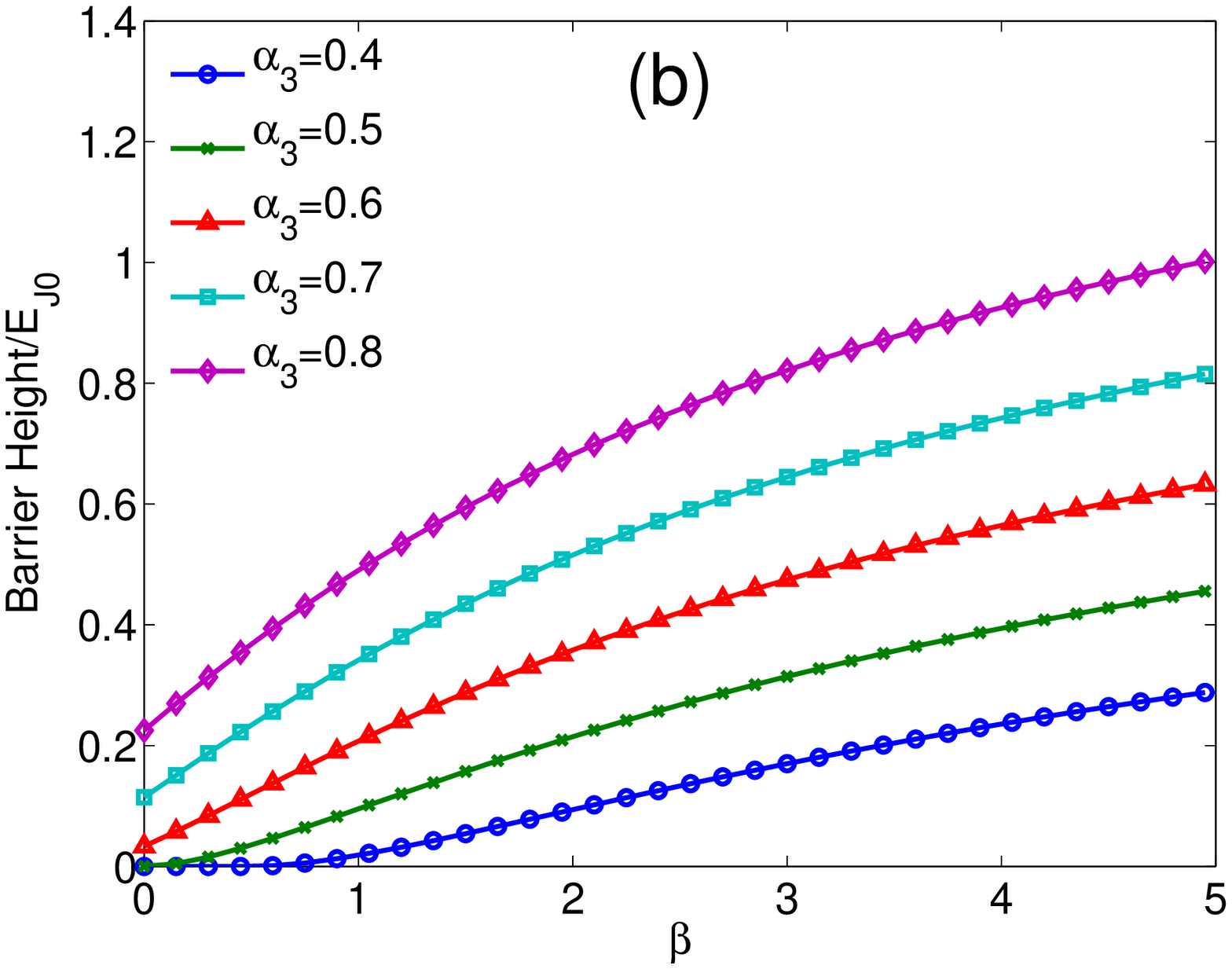}
\end{minipage}
\begin{minipage}[c]{0.3\textwidth}
\centering
\includegraphics[keepaspectratio=true,width=.95\textwidth]{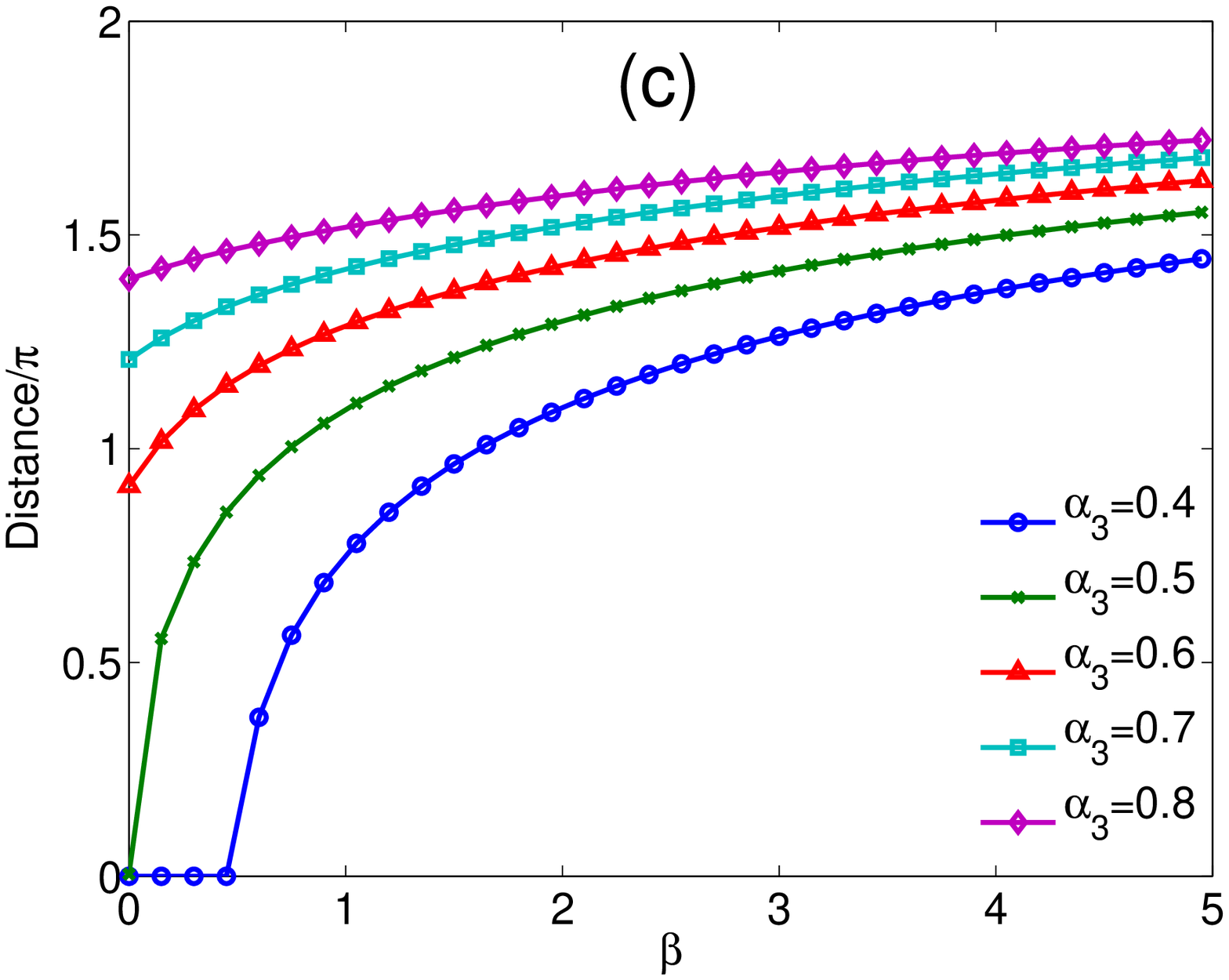}
\end{minipage}
\caption{\mycolor{}Features of the potential of the three-phase system in
Eq. (\ref{eq:ham_org}) against the reduced inductance $\beta$: (a) loop current in the
potential minima, (b) barrier height of the double-well structure,
(c) distance between two neighboring minima in the potential. Other
parameters are $\alpha_1=\alpha_2=1$, $f=0.5$ and $\alpha_3$
is shown in the legends.}
\label{fig:classical}
\end{figure}
To begin with, we first shortly discuss
the potential of the three-phase system at the degenerate point
$\phi_X=\pi$ or $f=\phi_X/2\pi=0.5$ with $\alpha_1=\alpha_2=1$ on the
coordinates $(\phi_1,\phi_2,\phi_3)$. %

\newcommand{\myPhi}[0]{ {\mathbb P} }

Figure \ref{fig:classical}(a) shows the magnitude of the loop current $I_{q}$
in the potential minima as a function of $\beta$, where different curves
correspond to different $\alpha_3$. When $\alpha_3<0.5$ and $\beta$ is small
enough, the bias flux cannot drive a persistent current in the loop, and a
zero-current point ${\myPhi}_{0}=(0,0,\pi)\mod2\pi$ achieves the potential
minimum $-\alpha_1-\alpha_2+\alpha_3$. As $\beta$ increases, the loop begins to
support a non-zero current. If $\alpha_3$ is smaller than a critical value
$\bar\alpha_3=\sqrt{2}/2$, which is obtained by
$I_{q}|_{\beta=0}=\alpha_3I_{C0}$, the increase of $\beta$ enlarges the loop
current to achieve the value $I_{C3}=\alpha_3I_{C0}$, the maximum current the
loop can afford. In other conditions, a larger inductance always suppresses the
loop current. As $\beta$ is large enough, e.g.  $\beta=5$, the inductance
aggressively erases the differences caused by $\alpha_3$ and damps $I_{q}$
into zero quickly, indicating the domination of the inductance $\beta$ in this
regime. A simple calculation shows that the phase of every junction tends to
be $0$ mod $2\pi$ and, therefore, the $\pi$-phase with which we bias the
circuit comes to drop on the loop inductance itself; for example, when
$\beta=5$ and $\alpha_3=0.8$, the inductance phase $\bar\phi$ reaches $0.60\pi$.
Actually, when $\beta$ is large enough, a tiny but non-zero current
approximately on $O(\Phi_0/2L)$ can make the inductance possess almost the
external $\pi$-phase bias and force the circuit to approach a possible global
potential minimum $-\sum_{k=1}^3\alpha_k$.


When the circuit begins to support a finite non-zero current in the potential
minima, its direction degeneracy yields that the potential minima such as
${\myPhi}_{+}=(\bar\phi^{+}_{1},\bar\phi^{+}_{2},\bar\phi^{+}_{3})$,
and ${\myPhi}_{-}=(-\bar\phi^{+}_{1},-\bar\phi^{+}_{2},2\pi-\bar\phi^{+}_{3})$, where
$\bar\phi^{+}_{1,2,3}$ belong to the interval of $(0,\pi)$, depart from each
other in pair while the zero-current point ${\myPhi}_{0}$ pins in the phase
space as a saddle point of the barrier, thus forming a double-well potential
structure. In this condition, Fig. \ref{fig:classical}(b) and (c) demonstrate the
barrier height defined by the potential difference between ${\myPhi
}_{+}$ and ${\myPhi}_{0}$ and the distance between ${\myPhi
}_{+}$ and ${\myPhi}_{-}$, respectively. It is found that a larger $\beta$ enhances the potential barrier and separates further the
well bottoms. Those numerical data verify an intuition that such a
non-negligible inductance suppresses the speed of switching the directions of the
loop current.  Consequently, in the quantum regime, these properties also
correspondingly weaken the interactions between the two persistent-current
states. See Ref. \onlinecite{Robertson2006} for a detailed discussion on the
three-phase system as well as its numerical method we utilized.\footnote{As we
know, in the new coordinates $(\phi,\theta_1,\theta_2)$, the three-phase potential
is of a crystal-slab style where the periodic variables $\theta_1$ and
$\theta_2$ represent the in-plane directions while the phase $\phi$
the normal one. Therefore, the solutions on $\theta_1$ and
$\theta_2$ should employ the Bloch theory widely used in the solid-state
physics.  Since no real crystal boundary condition such as the Born-Von Karman
boundary condition exists in the superconducting phase space, in a full
quantum description, we need not consider its energy band structures in the
reciprocal lattice space for most of cases but the size of the periodic cell in
the superconducting phase space should be minimized as we do in order to get the eigenenergy
at the original point of the first Brillouin zone. However,
Ref. \onlinecite{Robertson2006} selects a larger cell and takes account of the
tiny energy splittings according to the inter-cell interactions, which vanish
in our numerical results.}

To study the quantum behaviors of the flux qubit system,
the tight-binding model can be utilized in the first step,\cite{TJL1999,YAM2002,Ya2006} and the
Hamiltonian of the three-phase flux qubit with proper parameters can be
expanded approximately in its two flux states locating in two neighboring
potential minima respectively as
\begin{eqnarray}
    \label{eq:H_2level}
    \hat{H}&=&{\delta\Phi_X{I}_{p}}\sigma_z-\Delta\sigma_x,
\end{eqnarray}
where $\sigma_x$ and $\sigma_z$ are Pauli matrices, $\delta\Phi_X$ is the flux
deviation from the degeneracy point $\Phi_X=0.5\Phi_0$, $\Delta > 0$ is the
tunneling energy between two flux states and ${I}_{p}$ is the magnitude of
the characteristic current possessed by the flux states. Define the matrix element
$i_{mn}$ of the current operator $\hat{I}_{\phi}$ as $i_{mn} = \langle{m}|
\hat{I}_{\phi} |{n}\rangle$ for the $m$th and $n$th eigenstates $|m\rangle$
and $|n\rangle$ with eigenvalues $\varepsilon_{m}$ and $\varepsilon_{n}$,
respectively. At $f=0.5$, it yields
$2\Delta=\varepsilon_{1}-\varepsilon_{0}$ and $I_{p}$ can be calculated as the
magnitude of $i_{10}$. For comparison, the effective theory also
gives the corresponding results such as $2\tilde\Delta = \tilde\varepsilon_{1}
- \tilde\varepsilon_{0}$, $\tilde{i}_{mn}^{(2)}$, $\tilde{i}_{mn}^{(4)}$,
$\tilde{I}_{p}^{(2)}=|\tilde{i}_{01}^{(2)}|_{f=0.5}$ and
$\tilde{I}_{p}^{(4)}=|\tilde{i}_{01}^{(4)}|_{f=0.5}$, which we utilize the tilde
sign to symbolize in this section. The symbols $\tilde{i}_{mn}^{(2)}$ represent
the matrix elements of the effective but non-optimal current operator
$\tilde{I}_{\phi}^{(2)}$ and $\tilde{i}_{mn}^{(4)}$ are achieved by the optimal
effective current operator $\tilde{I}_{\phi}^{(4)}$. The numerical comparisons
are given as follows.

\begin{figure}
\centering
\begin{minipage}[c]{0.5\textwidth}
\centering
\includegraphics[keepaspectratio=true,width=1\textwidth]{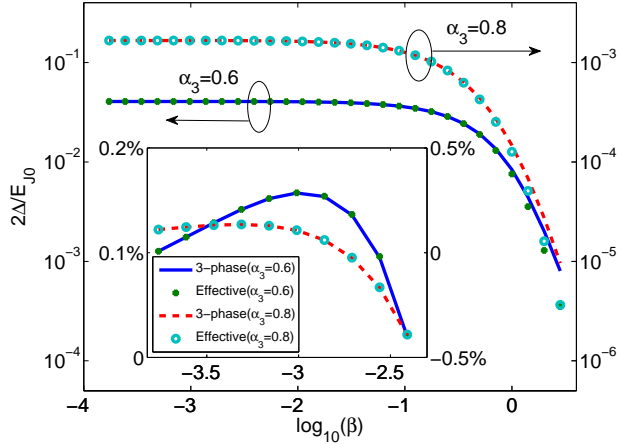}
\end{minipage}%
\caption{ \mycolor{}Tunnel splitting of the flux qubit{,} in units of $E_{J0}${,}
vs. the reduced inductance $\beta$. Parameter $\alpha_3$ is selected
to be equal to two typical values $0.6$ and $0.8$ while others are $\alpha_1=1$,
$\alpha_2=1$, $g=80$ and $f=0.5$. The inset with the same
symbols re-scales the range of $\beta$ and draws the percent changes of the numerical results to the corresponding values on $\beta=0$.}
\label{fig:delta}
\end{figure}

\begin{figure}
\centering
\begin{minipage}[c]{0.5\textwidth}
\centering
\includegraphics[keepaspectratio=true,width=1\textwidth]{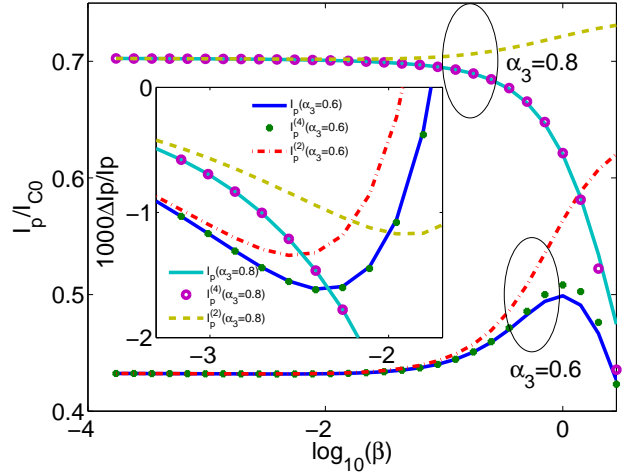}
\end{minipage}%
\caption{ \mycolor{}Current factors $I_{p}$, ${\tilde{I}_{p}^{(2)}}$ and $\tilde{I}_p^{(4)}$ in units of $I_{C0}$ as functions of
the reduced inductance $\beta$.
The inset plots the deviations of the currents from the corresponding values $I_{p}|_{\beta=0}$ on different $\alpha_3$ in per thousand. The parameter $\alpha_3$ is also chosen equal to $0.6$ and
$0.8$ while the others are $\alpha_1=1$, $\alpha_2=1$, $g=80$ and
$f=0.5$.} \label{fig:Ip}
\end{figure}
\begin{figure}
\begin{minipage}[c]{0.5\textwidth}
\begin{center}
\includegraphics[keepaspectratio=true,width=1\textwidth]{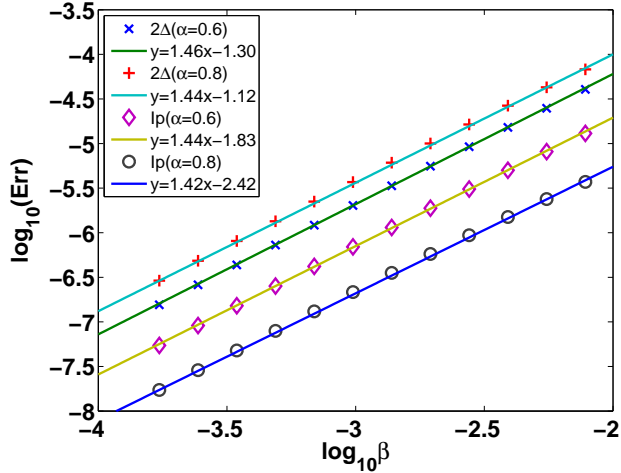}
\end{center}
\end{minipage}
\caption{ \mycolor{}Relative Errors $\frac{|\tilde\Delta-\Delta|}{|\Delta|_{\beta=0}}$ and
$\frac{|\tilde{I}^{(4)}_{p}-{I}_{p}|}{|{I}_{p}|_{\beta=0}}$ as functions of the
reduced inductance $\beta$.}
\label{fig:err}
\end{figure}
\begin{figure}
\begin{minipage}[c]{0.5\textwidth}
\centering
\includegraphics[keepaspectratio=true,width=1\textwidth]{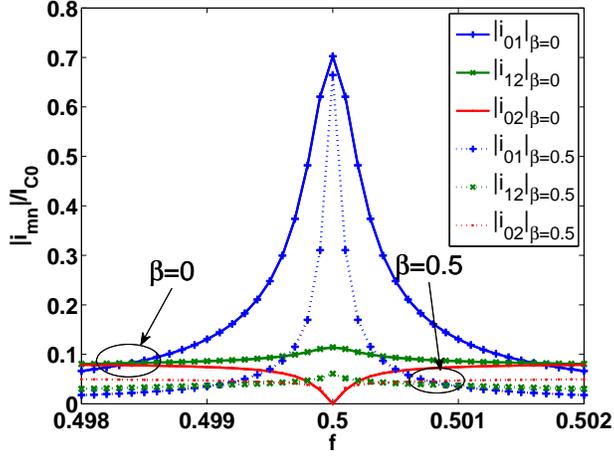}
\end{minipage}
\caption{ \mycolor{}Magnitudes of the current matrix elements $i_{mn}$ for
the three lowest levels, in units of $I_{C0}$, vs. the reduced
bias flux $f$. Two cases $\beta$=0 and $\beta$=0.5 are selected,
and other parameters are $\alpha_{1,2}=1$,$\alpha_3=0.8$ and
$g=80$.} \label{fig:i_mn}\end{figure}
\begin{figure}
\begin{minipage}[c]{0.5\textwidth}
\centering
\includegraphics[keepaspectratio=true,width=1\textwidth]{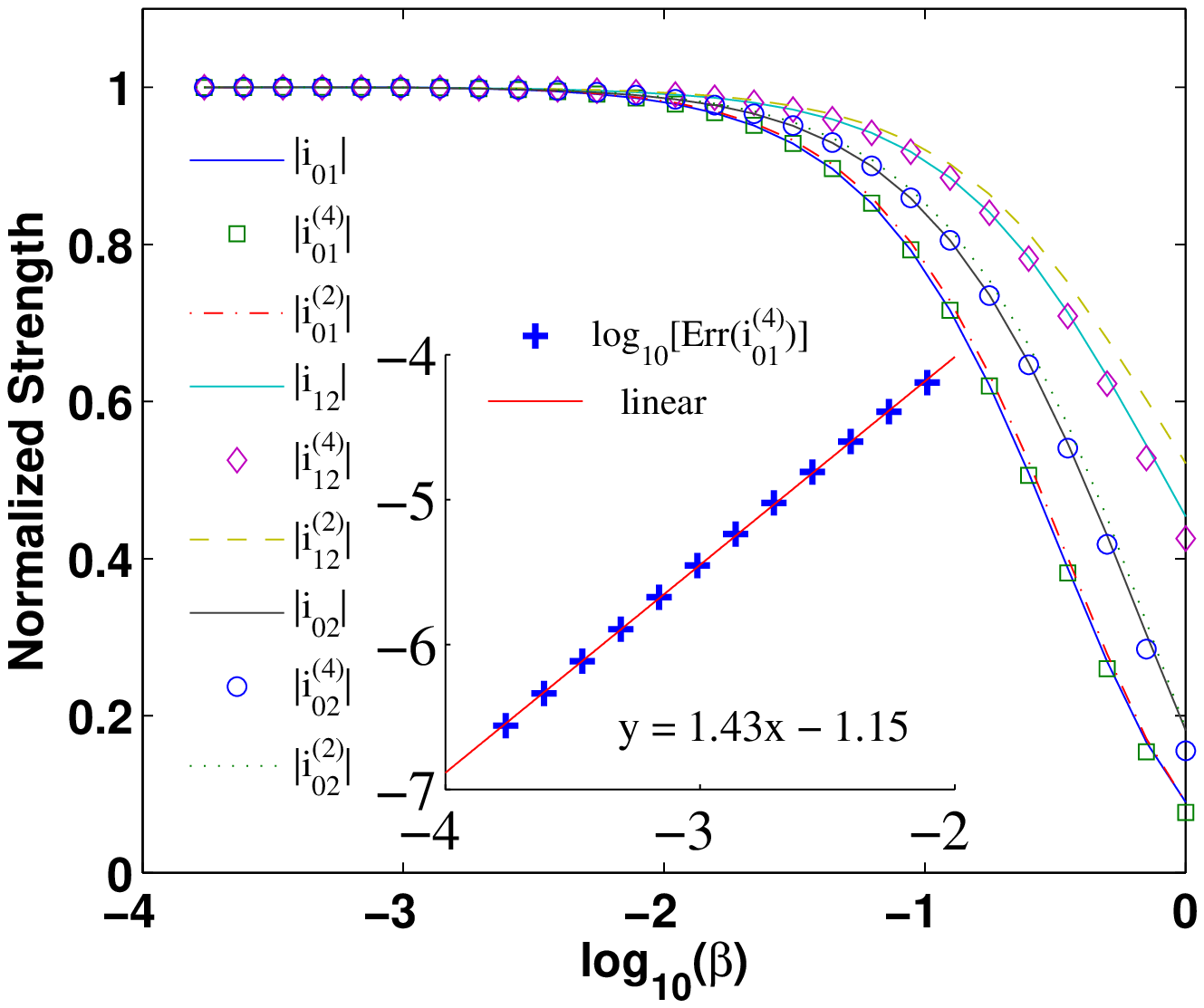}
\end{minipage}
\caption{ \mycolor{}Normalized $|i_{mn}|$, $|\tilde{i}_{mn}^{(2)}|$ and
$|\tilde{i}_{mn}^{(4)}|$ as functions of the reduced inductance $\beta$ when
$f=0.499$. The data $|{i}_{mn}|$, $|\tilde{i}_{mn}^{(2)}|$ and
$|\tilde{i}_{mn}^{(4)}|$ have been divided by the corresponding values
$|{i}_{mn}|_{\beta=0}$, and other parameters are
$\alpha_{1,2}=1$,$\alpha_3=0.8$ and $g=80$. The inset demonstrates a
typical error analysis.}
\label{fig:i_mn_norm}
\end{figure}

Figure \ref{fig:delta} plots $2\Delta$ and $2\tilde{\Delta}$ as functions of
$\beta$ based on $\alpha_3=0.6$ and $0.8$.  When $\beta$ is small enough, e.g.,
$\beta<0.1$, $2\Delta$ only deviates slightly from its inductance-free value.
Furthermore, when $\beta$ is re-scaled (see the inset), it is found that
$\Delta$ does not decrease monotonically but reaches its maximum value at
$\beta\simeq10^{-3}$. As mentioned above in the effective theory, the vacuum
fluctuations on $O(\beta^{1/2})$ brought by the LC-oscillator actually reduce
the effective sizes of the Josephson junctions, thus suppressing the barriers
and enhancing the interactions between these two flux states.  On the other
hand, the self-biased inductive effects like $-LI^2/2$ on $O(\beta)$  increase the barriers and slow down the current
direction switching speed. As a numerical $order$ prediction, we have those two
characteristic factors equal as $\gamma_1^2\simeq\beta$ and get a critical
value $\beta\sim10^{-3}$ agreeing with the data of the inset. As $\beta$
becomes larger, a clear tunnel rate damping means that the self-biased effects
grow up to a non-negligible level. When $\beta>1$, $2\Delta$ is
more than one order of magnitude smaller than its inductance-free value, and the effective result $2\tilde\Delta$ decays more excessively than $2\Delta$ does; in this
situation, a small $\Delta$ means that two flux states of the flux qubit
interact with each other weakly and slowly, rendering that the whole system fails to act as a useful qubit in
a larger $\alpha_3$ such as $\alpha_3=0.8$, but $\alpha_3=0.6$ only makes the
flux qubit slow down which may benefit the design on it with a large loop
inductance.  It is a pleasure that when the effective Hamiltonian on $O(\beta)$ fails to calculate
the inductive effects that involve higher excited levels of the oscillator, the
three-phase system with a set of traditional design parameters may no longer
perform as a good qubit.

To show the performances of the optimal effective current operator
$\tilde{I}_{\phi}^{(4)}$, Fig.
\ref{fig:Ip} depicts the numerical data of
${\tilde{I}_{p}^{(2)}}$, ${\tilde{I}_{p}^{(4)}}$ and $I_{p}$ vs. $\beta$ on the cases of $\alpha_3=0.6$ and
$\alpha_3=0.8$. There is no doubt that $\tilde{I}_{\phi}^{(4)}$
perfectly achieves the results in a high precision regardless of $\alpha_3$; even
when the inductance has a non-negligible size ($\beta\simeq1$),
it can also correctly predict the profiles of the $I_{p}-\beta$ curves.
These curves resemble their classical counterparts $I_q$ in Fig.
\ref{fig:classical}(a), which infers us that the shifts of the classical
potential minima introduced by a large inductance also take significant roles
in the quantum regime. Compared to $I_{p}$ and ${\tilde{I}_{p}^{(4)}}$, $\tilde{I}_p^{(2)}$
without full $O(\beta)$ corrections fails to describe $I_{p}$ when the
influences imposed by the inductance become notable, e.g. $\beta>0.01$, which
also emphasizes that the $O(\beta)$ effects dominate in this region. %
In fact, the inductive energy term
on $O(\beta)$ in $\tilde{H}^{(1)}_{eff}$ tends to make itself minimized averagely in a relatively large $\beta$
region which forces ${\tilde{I}_{p}^{(2)}}$ to rise too pronouncedly to
approximate the real value $I_{p}$. %
As mentioned above, the vacuum fluctuations of the LC oscillator bring in the
$O(\beta^{1/2})$ effects and, thus, reduce the effective sizes of the junctions.
Therefore, the currents are expected to decline when $\beta$ is small enough to
make the $O(\beta)$ effects negligible, which is also confirmed by the inset of Fig.
\ref{fig:Ip}. When $\alpha_3=0.8$, since the net $O(\beta)$ effects also
depress the currents ( see $I_{p}$ when $\beta>0.01$ ), $I_p$ and
${\tilde{I}_{p}^{(4)}}$ both monotonously decrease in the whole region. On the other hand,
lacking full $O(\beta)$ effects, %
${\tilde{I}_{p}^{(2)}}|_{\alpha_3=0.8}$ increases in the
large $\beta$ region, so there exists a minimum at $\beta\simeq0.01$ in the
corresponding curve when the $O(\beta^{1/2})$ and $O(\beta)$ effects strike
a balance. For $\alpha_3=0.6$, minima are also found to show the balances
between the opposite $O(\beta^{1/2})$ and $O(\beta)$ effects. Both of those
two types of minima support our previous conclusion that
$\beta\simeq10^{-3}\sim10^{-2}$ is the watershed to distinguish the region
dominated by the vacuum fluctuations. %
Figure \ref{fig:err} demonstrates the $\beta$-dependence of the errors which
$\tilde\Delta$ and $\tilde{I}_{p}^{(4)}$ bear. The linear fitting indicating
that these errors are approximately on $O(\beta^{1.5})$ sufficiently verifies our
analytic conclusions.

When a small magnitude of time-dependent flux is applied to drive the circuit,
the matrix elements $i_{mn}$ contribute to the strengths of the photon-assisted transition
rates, significant for the control of the circuit.\cite{YJL2005} We consider its
three lowest levels and plot the magnitudes of the matrix elements $i_{01,02,12}$  vs. the reduced flux bias $f$ in Fig. \ref{fig:i_mn}.
When $\beta$ increases to $\beta=0.5$, the $|i_{mn}|$-$f$ curves keep
the same shapes approximately except for that their line-widths are much narrower, meaning
that it is more difficult to control the circuit. The magnitudes shown in Fig.
\ref{fig:i_mn_norm}, as $f$ deviates slightly from $f=0.5$, are consistent with
our previous conclusion on the weakened interactions, and also indicate that
the effective current operator can accurately predict the results even when $\beta$ reaches one. The inset of Fig. \ref{fig:i_mn_norm}
also supports our order analysis.

To sum up, the optimal effective current operator describes the loop inductive
effects in good agreement with the three-phase full quantum predictions even when
the size of the inductance $L$ is comparable with the effective ones' of the junctions ($\beta\sim1$) and, consequently, the circuit may perform as a less useful qubit.

\section{conclusion} \label{sec:co}

In summary, an optimal effective two-phase current operator for 3jj flux qubit
has been obtained if considering the inductance of the circuit loop.  In our
classical analysis, we have utilized a source transformation to achieve the
current form for the inductance-free two-phase system. Then after
constructing the Hamiltonian for the three-phase system in the original phase space
$(\phi_1,\phi_2,\phi_3)$, we choose a reasonable linear transformation to
reformat the Hamiltonian in new variables, where the small inductance phase
$\phi$ is separated as a single coordinate, and find that the
system can be treated as the inductance-free flux qubit interacting with a high
frequency LC-oscillator. Under the condition that the energy of the
slow two-phase flux qubit system is small enough comparing to the
LC-oscillator's, an effective theory has been developed for physical variable
operators from the photon transition path method based on the BW expansion, which is also suitable for other
superconducting circuit types.  As an application for the relatively simple results which are still of high accuracy, the effective Hamiltonian is expanded to order $\beta$ and only give an error
on $O(\beta^{3/2})$.  Besides the direct expansion, enlightened
by the classical view on the circuit, we have also presented another simpler
method in the effective theory to achieve the explicit form of the optimal
current operator $\tilde{I}_{\phi}^{(4)}$ whose corresponding error is merely on
the order of $\beta^{3/2}$. Finally, we have verified that the optimal
effective operators perfectly describe the numerical properties of the three-phase
system.

\begin{acknowledgments}
Fruitful discussions with A. Maassen van den Brink, Jilin Wang, Tiefu Li, Peiyi
Chen, Zhibiao Hao, Yi Luo, Zhiping Yu and Zhijian Li are acknowledged. The
first author also expresses his great gratitude to Yufeng Li and Naijun Li for their
warm encouragement. This work is supported by the State Key Program for Basic
Research of China under Grant No. 2006CB921106 and 2006CB921801.
\end{acknowledgments}

\appendix
\section{order analysis}
\label{sec:order}
For $\hat{H}_{int}$ in Eq. (\ref{eq:H_int}), we define its expansion
 as
\begin{eqnarray}
\langle\Omega_m|\hat{H}_{int}|\Omega_n\rangle&=&
\sum_{k\ge1}\hat{\mathcal
V}_{m,n}^{(k)}\beta^{k/4},\label{eq:H_int_mn}
\end{eqnarray}
where $\hat{\mathcal V}_{m,n}^{(k)}$ are independent of $\beta$. According to
the optical selection rules, we have a $\beta^{1/2}$ (instead of
$\beta^{1/4}$) series expansion on $\hat{H}_{s,s_{1}}$ as
\begin{eqnarray}
\hat{H}_{s,s_{1}}&=&\beta^{|s-s_1|/4}\sum_{k\ge0}\hat{\mathcal
V}_{s,s_1}^{(|s-s_1|+2k)}\beta^{k/2}\sim{O}(\beta^{|s-s_1|/4}),
\label{eq:H_ss1_exp}
\end{eqnarray}
where the operator $\hat{\mathcal V}_{s,s}^{(0)}$ is an alias for the
Hamiltonian $\hat{H}_0$, and ``$\sim{O}(\beta^{|s-s_1|/4})$'' denotes %
that the dominant terms in the corresponding operator $\hat{H}_{s,s_{1}}$ are on ${O}(\beta^{|s-s_1|/4})$. %
The operator $\hat{H}_{s,s_{1}}$
combines the effects in proportion to $\beta^{|s-s_1|/4}$,
$\beta^{|s-s_1|/4+1/2}$,
$\beta^{|s-s_1|/4+1}$ and etc. Generally, the product
$\prod_{k=1}^{n}\hat{H}_{s_{k},s_{k-1}}$  $\sim$
$O(\beta^{(\sum_{k=1}^{n}|s_{k}-s_{k-1}|)/4})$ can also be expanded in a
$\beta^{1/2}$-series, where $n$ and $s_k$ are integers.

Operators $\hat{\mathcal H}_{s,s_1}$ introduced do not complicate the order
analysis on both $\tilde{H}(\varepsilon)$ and $\hat{\mathcal
G}_s(\varepsilon)$. The denominator $s\hbar\omega_{LC}$ is a
scale factor in proportion to $\beta^{-1/2}$, and $\varepsilon$ in
$\hat{\mathcal H}_{s,s}(\varepsilon)$ is dominated by $\varepsilon^{(0)}$ independent of
$\beta$. Therefore, one can obtain a typical term $\tilde{H}_{typ}(\varepsilon)$ in
$\tilde{H}(\varepsilon)$ as
\begin{eqnarray}
  \label{eq:H_typ}
    \tilde{H}_{typ}(\varepsilon)&=& \hat{H}_{0,s_0}\hat{\mathcal
    H}_{s_0,s_0}^{i_{0}}(\varepsilon)\hat{\mathcal H}_{s_0,s_{1}}\hat{\mathcal
    H}_{s_{1},s_{1}}^{i_{1}}(\varepsilon)\hat{\mathcal
    H}_{s_{1},s_2}...\hat{\mathcal
    H}_{s_n,s_n}^{i_n}(\varepsilon)\hat{\mathcal
    H}_{s_n,0}\\\nonumber&\sim&O(\beta^{\frac{n+1}{2}+\frac{1}{2}\Sigma_{k=0}^{n}i_{k}+\frac{1}{4}\Sigma_{k=1}^{n}|s_{k}-s_{k-1}|+\frac{s_0}{4}+\frac{s_n}{4}}),
\end{eqnarray}
where $n$, $s_{0}$, $\dots$, $s_{n}$, $i_0$, $\dots$ and $i_n$ are
integers, and $s_k\neq s_{k+1}$ for $k$ as an integer. The
$\varepsilon$-$\tilde{H}(\varepsilon)$ substitution can change
$\tilde{H}(\varepsilon)$ into a $\varepsilon$-independent one on a
specific order via using the first few largest $\varepsilon$-less terms
like $\hat{H}_{0}$ to replace
$\varepsilon$. Mathematically, with $n$ being an integer and $r=n/2$, we can
respectively expand the non-Hermitian Hamiltonian $\tilde{H}^{(r)}_{nh}$ to order
$\beta^{r}$ and the operator $\hat{\mathcal G}_s^{(r)}$ to order $\beta^{s/4+r}$ as
\begin{eqnarray}
    \label{eq:H_n_nh}
    \tilde{H}^{(r)}_{nh}&=&
    \tilde{H}^{(r)}_{nh}|_{\beta^0}+\tilde{H}^{(r)}_{nh}|_{\beta^{1/2}}+\tilde{H}^{(r)}_{nh}|_{\beta}+\dots+\tilde{H}^{(r)}_{nh}|_{\beta^{r}}+{\mathit
    O}(\beta^{r+\frac{1}{2}}), \\
    \label{eq:G_n_eff}
    \hat{\mathcal G}_s^{(r)}&=&
    \hat{\mathcal G}_s^{(r)}|_{\beta^{\frac{s}{4}+\frac{1}{2}}}+\hat{\mathcal
    G}_s^{(r)}|_{\beta^{\frac{s}{4}+\frac{2}{2}}}+\hat{\mathcal
    G}_s^{(r)}|_{\beta^{\frac{s}{4}+\frac{3}{2}}}+\dots\\\nonumber&&+\hat{\mathcal
    G}_s^{(r)}|_{\beta^{\frac{s}{4}+r}}+O(\beta^{\frac{s}{4}+\frac{1}{2}+r}),
\end{eqnarray}
where $\tilde{H}^{(r)}_{nh}|_{\beta^{k/2}}\propto\beta^{k/2}$ and
$\hat{\mathcal G}_s^{(r)}|_{\beta^{s/4+k/2}}\propto\beta^{s/4+k/2}$
for the integer $k$. %
With Eqs. (\ref{eq:H_n_nh}) and (\ref{eq:G_n_eff}), it can be verified correct
that the operator $\hat{\mathcal{G}}_{||}$ in Eq. (\ref{eq:G_n}) does not hold
any term in proportion to $\beta^{\frac{2k+1}{4}}$ for $k$ as an integer.
Therefore, the statement on $\tilde{H}_{eff}$ in Eq. (\ref{eq:eff_H_eq0}) can
be justified without any doubt. %
Since the effective states and their eigenenergy are determined by
the effective Hamiltonian $\tilde{H}_{eff}$, we also have
$\varepsilon^{(2k+1)}\equiv0$ and $|\varphi_{0}^{(2k+1)}\rangle\equiv0$ for $k$
as an integer in Eqs. (\ref{eq:eps_exp}) and (\ref{eq:phi_exp}), respectively.

\section{details of operator expansions}
\label{sec:op_exp}

In the following sections, $I_{C0}$ is utilized as the current unit, $\hbar=1$
and $e=1/2$. First of all, the operator $\hat{H}_{s,s_1}$ ( here $s$ may equate to $s_{1}$ ) in Eq.
(\ref{eq:Ham_mat}) is
\begin{eqnarray}
\label{eq:app_H_mn}
\hat{H}_{s,s_1}&=&
\hat{H}_0\delta_{s,s_1}+\langle\Omega_s|\left[\sum_{k=1}^{3}\alpha_{k}\cos\theta_{k}-\alpha_{k}\cos\left(\theta_{k}+\frac{\alpha_{ser}}{\alpha_k}\phi\right)\right]|\Omega_{s_1}\rangle
\\\nonumber
&=&\hat{H}_0\delta_{s,s_1}+\sum_{k=1}^3\alpha_k\cos\theta_k\delta_{s,s_1}-\frac{1}{2}\sum_{k=1}^3\left(\alpha_{k}e^{-\frac{\gamma_k^2}{2}+i\theta_k}\sum_{t=0}^{\min(s,s_1)}\frac{\sqrt{s!s_1!}(i\gamma_k)^{s+s_1-2t}}{t!(s-t)!(s_1-t)!}+c.c.\right),
\end{eqnarray}
where
$\gamma_k=\frac{\alpha_{ser}}{\alpha_k}\sqrt[4]{\frac{2\beta}{g\alpha_{ser}}},k=1,2,3$,
have been defined in Eq. (\ref{eq:gamma}); especially, we have
\begin{eqnarray}
\label{eq:H01}
\hat{H}_{0,1}&=&
\sum_{k=1}^3\alpha_{k}\gamma_ke^{-\frac{\gamma_k^2}{2}}\sin\theta_k,\\
\label{eq:H02}
\hat{H}_{0,2}&=&
\frac{\sqrt{2}}{2}\sum_{k=1}^3\alpha_{k}\gamma_k^2e^{-\frac{\gamma_k^2}{2}}\cos\theta_k.
\end{eqnarray}
Equation (\ref{eq:app_H_mn}) as an explicit expression is consistent with the
ones in our former order analysis such as Eq. (\ref{eq:H_ss1_exp}).
For $s=s_1$, the operator $\hat{H}_{s,s}$ holds its dominant term $\hat{H}_0$ independent of
$\beta$. Since $e^{-\frac{\gamma_k^2}{2}}=1+O(\beta^{1/2})$ and the largest
term among $(i\gamma_k)^{s+s_1-2t}$ in the sum is on $O(\beta^{|s-s_1|/4})$ when
$t$ equates to $\min(s,s_1)$, we have the dominant term of $\hat{H}_{s,s_1}$ on
$O(\beta^{|s-s_1|/4})$ for $s\neq s_1$. It is due to the optical selection rules
that the fluctuation factors $e^{-\frac{\gamma_k^2}{2}}$ as well as the sums about $(i\gamma_k)^{s+s_{1}-2t}$ can be expanded in a $\beta^{1/2}$-series. Thus, the
operator $\hat{H}_{s,s_1}$ is capable to be expanded
in the same way. Equation (\ref{eq:app_H_mn}) also yields
\begin{eqnarray}
  \label{eq:H00Hss}
  \hat{H}_{0,0}=\hat{H}_{s,s}+O(\beta^{1/2}),
\end{eqnarray}
where $s\neq0$, and the terms in proportion to $\beta^{1/4}$ miss due to the optical
selection rules.

We expand $\hat{\mathcal G}_{1,2,3}(\varepsilon)$ as follows,
\begin{eqnarray}
\label{eq:app_G_1}
\hat{\mathcal G}_1(\varepsilon)&=&\hat{\mathcal
H}_{1,0}+\hat{\mathcal H}_{1,1}(\varepsilon)\hat{\mathcal H}_{1,0}+\hat{\mathcal
H}_{1,1}^2(\varepsilon)\hat{\mathcal H}_{1,0}+\hat{\mathcal H}_{1,2}\hat{\mathcal
H}_{2,0}+O(\beta^{9/4})\\\nonumber
&=&
-\frac{\hat{H}_{1,0}}{\hbar\omega_{LC}}+\frac{(\hat{H}_{1,1}-\varepsilon)\hat{H}_{1,0}}{(\hbar\omega_{LC})^2}-\frac{(\hat{H}_{1,1}-\varepsilon)^2\hat{H}_{1,0}}{(\hbar\omega_{LC})^3}+\frac{\hat{H}_{1,2}\hat{H}_{2,0}}{2(\hbar\omega_{LC})^2}+O(\beta^{9/4})\\\nonumber
&=&
-\frac{\hat{H}_{1,0}}{\hbar\omega_{LC}}+\frac{\hat{H}_{1,1}\hat{H}_{1,0}-\hat{H}_{1,0}\hat{H}_{0,0}}{(\hbar\omega_{LC})^2}-\frac{\left[\hat{H}_{0},\left[\hat{H}_{0},\hat{H}_{1,0}\right]\right]}{(\hbar\omega_{LC})^3}\\\nonumber&&+\frac{\hat{H}_{1,2}\hat{H}_{2,0}}{2(\hbar\omega_{LC})^2}+O(\beta^{9/4}),\\
\label{eq:app_G_2}
\hat{\mathcal G}_2(\varepsilon)&=&\hat{\mathcal
H}_{2,0}+\hat{\mathcal H}_{2,2}(\varepsilon)\hat{\mathcal H}_{2,0}+\hat{\mathcal H}_{2,1}\hat{\mathcal
H}_{1,0}+O(\beta^{8/4})\\\nonumber
&=&
-\frac{\hat{H}_{2,0}}{2\hbar\omega_{LC}}+\frac{(\hat{H}_{2,2}-\varepsilon)\hat{H}_{2,0}}{4(\hbar\omega_{LC})^2}+\frac{\hat{H}_{2,1}\hat{H}_{1,0}}{2(\hbar\omega_{LC})^2}+O(\beta^{8/4})\\\nonumber
&=&
-\frac{\hat{H}_{2,0}}{2\hbar\omega_{LC}}+\frac{\left[\hat{H}_{0},\hat{H}_{2,0}\right]}{4(\hbar\omega_{LC})^2}+\frac{\hat{H}_{2,1}\hat{H}_{1,0}}{2(\hbar\omega_{LC})^2}+O(\beta^{8/4}),\\
\label{eq:app_G_3}
\hat{\mathcal G}_3(\varepsilon)&=&\hat{\mathcal
H}_{3,0}+O(\beta^{7/4})=-\frac{\hat{H}_{3,0}}{3\hbar\omega_{LC}}+O(\beta^{7/4}),
\end{eqnarray}
where since from the effective Hamiltonian we know that
\begin{eqnarray}
\tilde{H}(\varepsilon)=\hat{H}_{0}+O(\beta^{1/2})=\hat{H}_{0,0}+O(\beta),
\end{eqnarray}
we utilize the $\varepsilon$-$\tilde{H}(\varepsilon)$ substitutions on
specific orders of $\beta$ as follows:
\begin{eqnarray}
\frac{\varepsilon\hat{H}_{1,0}}{(\hbar\omega_{LC})^2}|\varphi_0\rangle&=&
\frac{\hat{H}_{1,0}\tilde{H}(\varepsilon)}{(\hbar\omega_{LC})^2}|\varphi_0\rangle=\frac{\hat{H}_{1,0}\hat{H}_{0,0}}{(\hbar\omega_{LC})^2}|\varphi_0\rangle+O(\beta^{9/4}),\label{eq:h10_o2}\\
\frac{(\hat{H}_{1,1}-\varepsilon)^2\hat{H}_{1,0}}{(\hbar\omega_{LC})^3}|\varphi_0\rangle&=&
\frac{(\hat{H}_{1,1}^2+\varepsilon^2-2\hat{H}_{1,1}\varepsilon)\hat{H}_{1,0}}{(\hbar\omega_{LC})^3}|\varphi_0\rangle\\\nonumber
&=&
\frac{\hat{H}_{1,1}^2\hat{H}_{1,0}+\hat{H}_{1,0}\tilde{H}(\varepsilon)^2-2\hat{H}_{1,1}\hat{H}_{1,0}\tilde{H}(\varepsilon)}{(\hbar\omega_{LC})^3}|\varphi_0\rangle\\\nonumber
&=&
\frac{\left[\hat{H}_{0},\left[\hat{H}_{0},\hat{H}_{1,0}\right]\right]}{(\hbar\omega_{LC})^3}|\varphi_0\rangle+O(\beta^{9/4}),\label{eq:h11_2}\\
\frac{\varepsilon\hat{H}_{2,0}}{4(\hbar\omega_{LC})^2}|\varphi_0\rangle&=&
\frac{\hat{H}_{2,0}\tilde{H}(\varepsilon)}{4(\hbar\omega_{LC})^2}|\varphi_0\rangle=\frac{\hat{H}_{2,0}\hat{H}_{0}}{4(\hbar\omega_{LC})^2}|\varphi_0\rangle+O(\beta^{8/4}).\label{eq:h2}
\end{eqnarray}

\section{solution of generalized eigen-problem}
\label{sec:sol}

For the generalized eigen-problem, due to the perturbations, the positive and
Hermitian operator $\hat{R}$ suggests that Eq. (\ref{eq:H_ord_6_4_eq}) can be converted to an
eigen-problem as
\begin{eqnarray}
  \label{eq:H_ord_6_4_herm}
  \hat{R}^{-\frac{1}{2}}\tilde{H}^{(3/2)}_{L}\hat{R}^{-\frac{1}{2}}\left(\hat{R}^{\frac{1}{2}}|\varphi_{0}\rangle\right)&=&\varepsilon\left(\hat{R}^{\frac{1}{2}}|\varphi_{0}\rangle\right)+O(\beta^{2}),
\end{eqnarray}
where the initial ``$+O(\beta^{7/4})$'' has been improved to ``$+O(\beta^{2})$'' due to the previous discussions on the optical selection rules.
Expanding $\hat{R}$ to order $\beta^{3/2}$ in Eq. (\ref{eq:H_ord_6_4_herm})
yields
\begin{eqnarray}
  \label{eq:H_ord_6_4_herm_1}
  \tilde{H}^{(3/2)}_{eff}|\varphi_{eff}^{(3/2)}\rangle&=&\varepsilon|\varphi_{eff}^{(3/2)}\rangle+O(\beta^{2}),
\end{eqnarray}
where an effective Hamiltonian $\tilde{H}^{(3/2)}_{eff}$ independent of
$\varepsilon$ reads
\begin{eqnarray}
  \label{eq:H_ord_6_4_eff}
  \tilde{H}^{(3/2)}_{eff} &=& \hat{H}_{0,0}-\frac{\hat{H}_{0,1}\hat{H}_{1,0}}{\hbar\omega_{LC}}-\frac{\hat{H}_{0,2}\hat{H}_{2,0}}{2\hbar\omega_{LC}}+\frac{\hat{H}_{0,1}\hat{H}_{1,1}\hat{H}_{1,0}}{(\hbar\omega_{LC})^2}\\\nonumber&&-\frac{1}{2}\frac{\hat{H}_{0,1}\hat{H}_{1,0}\hat{H}_{0,0}+\hat{H}_{0,0}\hat{H}_{0,1}\hat{H}_{1,0}}{(\hbar\omega_{LC})^2},
\end{eqnarray}
and the effective eigenstate $|\varphi_{eff}^{(3/2)}\rangle$ is
\begin{eqnarray}
  \label{eq:H_ord_6_4_st}
  |\varphi_{eff}^{(3/2)}\rangle&=& \left(1+\frac{1}{2}\frac{\hat{H}_{0,1}\hat{H}_{1,0}}{(\hbar\omega_{LC})^2}\right)|\varphi_{0}\rangle.
\end{eqnarray}
The effective Hamiltonian $\tilde{H}^{(3/2)}_{eff}$ is Hermitian since the
transformation $\left(\hat{R}^{-\frac{1}{2}}\cdot\hat{R}^{-\frac{1}{2}}\right)$
does not alter the Hermiticity of $\tilde{H}^{(3/2)}_{L}$. One
remarkable thing is that $|\varphi_{eff}^{(3/2)}\rangle$ is naturally normalized
on $O(\beta^{3/2})$. %
Since the eigenstate $|\varphi\rangle$ is normalized as
$\langle\varphi|\varphi\rangle=1$, we expand it to order $\beta^{3/2}$ and
have
\begin{eqnarray}
\label{eq:eff_state_norm}
  1=\langle\varphi|\varphi\rangle=\langle\varphi_{eff}^{(3/2)}|\varphi_{eff}^{(3/2)}\rangle+O(\beta^{2}).
\end{eqnarray}
In sum, Eq. (\ref{eq:H_ord_6_4_herm_1}) is consistent with Eq.
(\ref{eq:H_eff_6_4}) as an eigen-problem which covers the eigenstates in the manifold $\mathcal{M}_{0}$ for the whole
three-phase system on $O(\beta^{3/2})$.

\section{proofs of Hermiticity of $\tilde{H}_{eff}$}
\label{sec:pf_herm}

First, let us calculate the value of the operator $\hat{D}$
\begin{eqnarray}
  \label{eq:pf_1}
  \hat{D}=\hat{\mathcal{G}}_{||}^2\tilde{H}_{nh}-\tilde{H}_{nh}^{\dag}\hat{\mathcal{G}}_{||}^{2},
\end{eqnarray}
According to Eqs. (\ref{eq:H_nh}) and (\ref{eq:orth_2p}), applying
$\langle\psi_0|\cdot|\varphi_0\rangle$ to Eq. (\ref{eq:pf_1}) yields
\begin{eqnarray}
  \label{eq:pf_2}
  \langle\psi_0|\hat{D}|\varphi_0\rangle&=& \langle\psi_0|\hat{\mathcal{G}}_{||}^2\tilde{H}_{nh}|\varphi_0\rangle-\langle\psi_0|\tilde{H}_{nh}^{\dag}\hat{\mathcal{G}}_{||}^{2}|\varphi_0\rangle\\\nonumber
  &=&(\varepsilon_\varphi-\varepsilon_\psi)\delta_{\psi,\varphi}\\\nonumber
  &\equiv &0.
\end{eqnarray}
Assuming the dominant term $\hat{D}^{(k)}$ of $\hat{D}$ is proportional to
$\beta^{k/4}$ with $k$ being an integer, we can expand Eq. (\ref{eq:pf_2}) to
order $\beta^{k/4}$ as
\begin{eqnarray}
  \label{eq:pf_3}
  \langle\psi_0^{(0)}|\hat{D}^{(k)}|\varphi_0^{(0)}\rangle\equiv0.
\end{eqnarray}
As the projected components $|\psi_0^{(0)}\rangle$ and $|\varphi_0^{(0)}\rangle$ are arbitrary
eigenstates of the unperturbed Hamiltonian $\hat{H}_{0}$, it yields
\begin{eqnarray}
  \label{eq:pf_4}
  \hat{D}^{(k)}\equiv0,
\end{eqnarray}
and, thus, %
\begin{eqnarray}
  \label{eq:pf_5}
  \hat{D}\equiv0,
\end{eqnarray}
for $k$ being arbitrary. It follows that
\begin{eqnarray}
  \label{eq:pf_6}
  \hat{\mathcal{G}}_{||}^2\tilde{H}_{nh}=\tilde{H}_{nh}^{\dag}\hat{\mathcal{G}}_{||}^{2}.
\end{eqnarray}
Finally, we achieve that
\begin{eqnarray}
  \label{eq:pf_7}
  \tilde{H}_{eff}&=&\hat{\mathcal{G}}_{||}\tilde{H}_{nh}\hat{\mathcal{G}}_{||}^{-1}\\\nonumber
  &=& \hat{\mathcal{G}}_{||}^{-1}\tilde{H}_{nh}^{\dag}\hat{\mathcal{G}}_{||}\\\nonumber
  &=& \tilde{H}_{eff}^{\dag}.
\end{eqnarray}

\section{calculations on effective operators}
\label{sec:de}

For the charge operator
\begin{eqnarray}
\hat{Q}_\phi&=&
i\sqrt[4]{\frac{\alpha_{ser}g}{2\beta}}\frac{\hat{a}^{\dag}-\hat{a}}{2},
\end{eqnarray}
we have its effective operator on $O(\beta)$ as
\begin{eqnarray}
  \tilde{Q}_\phi &=&i\sqrt[4]{\frac{\alpha_{ser}g}{2\beta}}\frac{\hat{\mathcal G}_1^{\dag}-\hat{\mathcal G}_1}{2}+O(\beta^{3/2})
  \\\nonumber&=&
  \frac{i\beta\alpha_{ser}g}{8}{\left[\tilde{I}_{\phi}^{(2)},\hat{H}_{0}\right]}+O(\beta^{3/2}).
\end{eqnarray}
With $|\varphi\rangle$, $|\psi\rangle$, $\varepsilon_\varphi$ and
$\varepsilon_\psi$ being defined in Sec. \ref{sec:ph:ar}, we apply
$\langle\psi|\cdot|\varphi\rangle$ to
$\dot{Q}_{\phi}$ and expand it on $O(\beta)$ as
\begin{eqnarray}
  \label{eq:H_Q_sigma_states}
  \langle\psi|\dot{Q}_{\phi}|\varphi\rangle &=& \langle\psi|\left[\hat{H}_{tr},i\hat{Q}_\phi\right]|\varphi\rangle\\\nonumber
  &=& i(\varepsilon_\psi-\varepsilon_\varphi)\langle\psi|\hat{Q}_\phi|\varphi\rangle\\\nonumber
  &=& \frac{-\beta\alpha_{ser}g(\varepsilon_\psi-\varepsilon_\varphi)}{8}\langle\psi_0|{\left[\tilde{I}_{\phi}^{(2)},\hat{H}_{0}\right]}|\varphi_0\rangle+O(\beta^{3/2}) \\\nonumber
  &=&\left(\frac{\varepsilon_\psi-\varepsilon_\varphi}{\hbar\omega_{LC}}\right)^2\langle\psi_0|\tilde{I}_{\phi}^{(2)}|\varphi_0\rangle+O(\beta^{3/2})\\\nonumber
  &=&\langle\psi_0|\left(\frac{\varepsilon_\psi^2\tilde{I}_{\phi}^{(2)}+\tilde{I}_{\phi}^{(2)}\varepsilon_\varphi^2-2\varepsilon_\psi\tilde{I}_{\phi}^{(2)}\varepsilon_\varphi}{\left(\hbar\omega_{LC}\right)^2}\right)|\varphi_0\rangle+O(\beta^{3/2})\\\nonumber
  &=&\langle\psi_0|\left(\frac{\hat{H}_{0}^2\tilde{I}_{\phi}^{(2)}+\tilde{I}_{\phi}^{(2)}\hat{H}_{0}^2-2\hat{H}_{0}\tilde{I}_{\phi}^{(2)}\hat{H}_{0}}{\left(\hbar\omega_{LC}\right)^2}\right)|\varphi_0\rangle+O(\beta^{3/2}),
\end{eqnarray}
where we utilize
\begin{eqnarray}
\label{eq:H0_e1}
\hat{H}_0|\varphi_0\rangle=\varepsilon_\varphi|\varphi_0\rangle+O(\beta^{1/2}),\\
\label{eq:H0_e2}
\hat{H}_0|\psi_0\rangle=\varepsilon_\psi|\psi_0\rangle+O(\beta^{1/2}).
\end{eqnarray}
Therefore, we have
\begin{eqnarray}
\dot{Q}_{\phi}^{eff}&=&\frac{\hat{H}_{0}^2\tilde{I}_{\phi}^{(2)}+\tilde{I}_{\phi}^{(2)}\hat{H}_{0}^2-2\hat{H}_{0}\tilde{I}_{\phi}^{(2)}\hat{H}_{0}}{\left(\hbar\omega_{LC}\right)^2}+O(\beta^{3/2}),
\label{eq:HQ_res}
\end{eqnarray}
which is the same as Eq. (\ref{eq:dQ_eff}).

\end{document}